\documentclass[prd,showpacs,twocolumn,nopacs,nofootinbib]{revtex4}
\usepackage{mathrsfs}
\usepackage{yfonts}
\usepackage{tipa}
\usepackage{bm}
\usepackage{bbm}
\usepackage{amsmath}
\usepackage{amssymb}
\usepackage{amsthm}
\usepackage{amsfonts}
\usepackage{mathtools}
\usepackage{latexsym}
\usepackage{relsize}
\usepackage[greek,english]{babel}
\usepackage{booktabs}
\usepackage[iso-8859-7]{inputenc}
\usepackage{tikz}
\usepackage{array}

\newcommand{\bea}{\begin{eqnarray}}
\newcommand{\eea}{\end{eqnarray}}
\newcommand{\be}{\begin{equation}}
\newcommand{\ee}{\end{equation}}

\begin{document}

\title{Palatini formulation of $f(R,T)$ gravity theory, and its cosmological
implications}
\author{Jimin Wu}
\email{ace227698890999@126.com}
\affiliation{School of Physics, Sun Yat-Sen University, Guangzhou 510275, People's
Republic of China}
\author{Guangjie Li}
\email{ligj6@icloud.com}
\affiliation{School of Physics, Sun Yat-Sen University, Guangzhou 510275, People's
Republic of China}
\author{Tiberiu Harko}
\email{t.harko@ucl.ac.uk}
\affiliation{Department of Physics, Babes-Bolyai University, Kogalniceanu Street,
Cluj-Napoca 400084, Romania}
\affiliation{School of Physics, Sun Yat-Sen University, Guangzhou 510275, People's
Republic of China}
\affiliation{Department of Mathematics, University College London, Gower Street, London
WC1E 6BT, United Kingdom}
\author{Shi-Dong Liang}
\affiliation{School of Physics, Sun Yat-Sen University, Guangzhou 510275, People's
Republic of China}
\affiliation{State Key Laboratory of Optoelectronic Material and Technology, and
Guangdong \\
Province Key Laboratory of Display Material and Technology, Guangzhou,
People's Republic of China}
\email{stslsd@mail.sysu.edu.cn}

\begin{abstract}
We consider the Palatini formulation of $f(R,T)$ gravity theory, in
which a nonminimal coupling between the Ricci scalar and the trace of the
energy-momentum tensor is introduced, by considering the metric and the
affine connection as independent field variables. The field equations and
the equations of motion for massive test particles are derived, and we show
that the independent connection can be expressed as the Levi-Civita
connection of an auxiliary, energy-momentum trace dependent metric,
related to the physical metric by a conformal transformation. Similarly to
the metric case, the field equations impose the non-conservation of the
energy-momentum tensor. We obtain the explicit form of the equations of
motion for massive test particles in the case of a perfect fluid, and the
expression of the extra-force, which is identical to the one obtained in the
metric case. The thermodynamic interpretation of the theory is also briefly
discussed. We investigate in detail the cosmological implications of the
theory, and we obtain the generalized Friedmann equations of the $f(R,T)$
gravity in the Palatini formulation. Cosmological models with Lagrangians of
the type $f=R-\alpha ^2/R+g(T)$ and $f=R+\alpha ^2R^2+g(T)$ are
investigated. These models lead to evolution equations
whose solutions describe accelerating Universes at late times.
\end{abstract}

\maketitle
\tableofcontents

\section{Introduction}

The observational discovery of the recent acceleration of the Universe \cite%
{1n,2n,3n,4n,acc} has raised the fundamental theoretical problem if general
relativity, in its standard formulation, can fully account for all the
observed phenomena at both galactic and extra-galactic scales. The simplest
theoretical explanation for the observed cosmological dynamics consists in
slightly modifying the Einstein field equations, by adding to it a
cosmological constant $\Lambda $ \cite{Wein}. Together with the assumption
of the existence of another mysterious component of the Universe, called
dark matter \cite{dm1,dm2}, assumed to be cold and pressureless, the
Einstein gravitational field equations with the cosmological constant
included, can give an excellent fit to all observed data, thus leading to
the formulation of the standard cosmological paradigm of our present days,
called the $\Lambda $ Cold Dark Matter ($\Lambda $ CDM) model. However,
despite its apparent simplicity and naturalness, the introduction of the
cosmological constant raises a number of important theoretical and
observational question for which no convincing answers have been provided so
far. The $\Lambda $CDM model can fit the observational data at a high level
of precision, it is a very simple theoretical approach, it is easy to use in
practice, but up to now no fundamental theory can explain it. Why is the
cosmological constant so small? Why it is so fine-tuned? And why the
Universe began to accelerate only recently? And, after all, a cosmological
constant would be really necessary to explain all observations?

From a theoretical point of view two possible answers to the questions
raised by the observation of the recent acceleration of the Universe can be
formulated. The first, called the \textit{dark energy approach}, assumes
that the Universe is filled by a mysterious and unknown component, called
dark energy \cite{PeRa03, Pa03,8,Tsu}, which is fully responsible for the
acceleration of the Universe, as well as for its mass-energy balance. The
cosmological constant $\Lambda$ corresponds to a particular phase of the
dynamical dark energy (ground state of a potential, let's say), and the
recent de Sitter phase may prove to be just an attractor of the dynamical
system describing the cosmological evolution. A second approach, the \textit{%
dark gravity approach}, assumes the alternative possibility that at large
scales the gravitational force may have a very different behavior as
compared to the one suggested by standard general relativity. In the general
relativistic description of gravity, the starting point is the
Hilbert-Einstein action, which can be written down as $S=\int{\
\left(R/2\kappa ^2+L_m\right)\sqrt{-g} \thickspace
\mathrm{d}^4 x}$, where $R$ is the Ricci scalar, $\kappa $ is the
gravitational coupling constant, and $L_m$ is the matter Lagrangian,
respectively. Hence in dark gravity type theories for a full understanding
of the gravitational interaction a generalization of the Hilbert-Einstein
action is necessary.

There are (at least) two possibilities to construct dark gravity theories.
The first is based on the modification of the geometric part of the
Hilbert-Einstein Lagrangian only. An example of such an approach is the $%
f(R) $ gravity theory, introduced in \cite{Bu,Ba}, and in which the
geometric part of the action is generalized so that it becomes an arbitrary
function $f(R)$ of the Ricci scalar. Hence in $f(R)$ gravity the total
Hilbert-Einstein action can be written as $S=\int{\left(f(R)/2\kappa
^2+L_m\right)\sqrt{-g} \thickspace
\mathrm{d}^4 x}$. The recent cosmological observations can be satisfactorily
explained in the $f(R)$ theory, and a solution of the dark matter problem,
interpreted as a geometric effect in the framework of the theory, can also
be obtained \cite{dm3}. For reviews and in depth discussions of $f(R)$ and
other modified gravity theories see \cite{r1,r2,r3,r4,rn, r5,r6,r7,r8}.

A second avenue for the construction of the dark gravity theories consists
in looking for maximal extensions of the Hilbert-Einstein action, in which
the matter Lagrangian $L_m$ plays an equally important role as the Ricci
scalar. Hence in this more general approach one modifies both the geometric
and the matter terms in the Hilbert-Einstein action, thus allowing a
coupling between matter and geometry \cite{Bert,Harkof}. The first
possibility for such a coupling is to replace the gravitational action by an
arbitrary function of the Ricci scalar and the matter Lagrangian $L_m$, thus
obtaining the so-called $f\left(R,L_m\right)$ class of modified gravity
theories \cite{fRLm}. This class of theories has the potential of explaining
the recent acceleration of the Universe without the need of the cosmological
constant, and can give some new insights into the dark matter problem, and
on the nature of the gravitational motion. The cosmological and physical
implications of this theory have been intensively investigated \cite%
{fRLm1,fRLm2,fRLm3,fRLm4,fRLm5,fRLm6,fRLm7, fRLm8,fRLm9,fRLm10,fRLm11}. For
a review of the $f\left(R,L_m\right)$ type theories see \cite{fRLm12}.

A second extension of the Hilbert-Einstein action can be obtained by
assuming that the gravitational field couples to the trace $T$ of the
energy-momentum tensor of the matter. This assumption leads to the $f(R,T)$
class of gravitational theories \cite{Harkof}. $f(R,T)$ theory may give some
hints for the existence of an effective classical description of the quantum
properties of gravity. As pointed out in \cite{fRT1}, by using a
nonperturbative approach for the quantization of the metric, proposed \cite%
{flu1,flu2,flu3}, as a consequence of the quantum fluctuations of the
metric, a particular type of $f(R, T )$ gravity naturally emerges, with the
Lagrangian given by $L =\left[(1 - \alpha )R/2\kappa^2 + \left(L_m - \alpha
T /2\right)\right]\sqrt{-g}$, where $\alpha $ is a constant. This
interesting theoretical result suggests that a deep connection may exist
between the quantum field theoretical description of gravity, which
naturally involves particle production in the gravitational field, and the
corresponding effective classical description of the $f(R,T)$ gravity theory
\cite{Liu}. The astrophysical and cosmological implications of $f(R,T)$
gravity theory were investigated in \cite{Ho0,Ha14,Ho1,fRT2, fRT2a, fRT3,
fRT4,fRT5,fRT6,fRT7,You1,You2,You3,fRT8,fRT9,fRT10,
fRT11,fRT12,fRT13,fRT14,fRT15,fRT16,fRT17,fRT18, fRT19,Shab1,Shab2,You4}.

Einstein's theory of general relativity can be obtained by starting from two
different theoretical approaches, called the metric and the Palatini
formalism, respectively, the later being introduced by Albert Einstein \cite%
{Ein1,Ein2,Ein3}. In the Palatini variational approach one takes as
independent field variables not only the ten components $g_{\mu \nu }$ of
the metric tensor, but also the components of the affine connection $\Gamma
_{\,\beta \mu }^{\alpha }$, without assuming, a priori, the form of the
dependence of the connection on the metric tensor, and its derivatives \cite%
{Olmo, Olmo1}. When applied to the Hilbert-Einstein action, these two
approaches lead to the same gravitational field equations. Moreover, the
Palatini formalism also provides the explicit form of the symmetric
connection as determined by the derivatives of the metric tensor. However,
in $f(R)$ modified gravity, as well as in other modified theories of
gravity, this does not happen anymore. In fact it turns out that the
gravitational field equations obtained by using the metric approach are
generally different from those obtained by using the Palatini variation \cite%
{Olmo,Olmo1}. An important difference is related to the order of the field
equations. The metric formulation usually leads to higher-order derivative
field equations, while in the Palatini formalism the obtained gravitational
field equations are always second order partial differential equations. A
number of new algebraic relations also do appear in the Palatini variational
formulation, which describe the subtle relation between the matter fields
and the affine connection, which can be determined from a set of equations
that couples it not only to the metric, but also to the matter fields. The
astrophysical and cosmological implications of the Palatini formulation of $%
f(R)$ gravity have been also intensively investigated \cite%
{PR1,PR2,PR3,PR4,PR5}.

Based on a hybrid combination of the metric and Palatini mathematical
formalisms, an extension of the $f(R)$ gravity theory was proposed in \cite%
{h1}, and was used to construct a new type of gravitational Lagrangian \cite%
{h1,h2}. A simple example of such hybrid metric-Palatini theory can be
constructed by adopting for the gravitational Lagrangian the expression $R +
f \left(R\left(g,\tilde{\Gamma}\right)\right)$, where $R\left(g,\tilde{\Gamma%
}\right)$ is the Palatini scalar curvature. A similar formalism that
interpolate between the metric and Palatini regimes was proposed in \cite%
{Ko1,Ko2} for the study of $f(R)$ type theories. This approach is called the
C-theory. A generalization of the hybrid metric-Palatini gravity was
introduced in \cite{B1}.

Despite the intensive investigations of the theoretical and observational
aspects of the modified gravity theories with geometry-matter coupling,
their Palatini formulation and properties have attracted considerably
smaller attention. The Palatini formulation of the linear $%
f\left(R,L_m\right)$ gravity was introduced in \cite{H-Pl}, where the field
equations and the equations of motion for massive test particles were
derived. The independent connection can be expressed as the Levi-Civita
connection of an auxiliary, matter Lagrangian dependent metric, which is
related to the physical metric by means of a conformal transformation.
Similarly to the metric case, the field equations impose the
non-conservation of the energy-momentum tensor. The study of Palatini
formulation of $f(R,T)$ gravity was initiated in \cite{Pal1}. Analogously to
its metric counterpart, the field equations impose of the $f(R,T)$ gravity
in the Palatini formulation implies the non-conservation of the
energy-momentum tensor, which leads to non-geodesic motion, and to the
appearance of an extra force.

It is the purpose of the present paper to derive the gravitational field
equations of the generalized $f(R,T)$ type gravity models, with non-minimal
coupling between matter, described by the trace of the energy-momentum
tensor, and geometry, characterized by the Ricci scalar By taking separately
two independent variations of the gravitational action with respect to the
metric and the connection, respectively, we obtain the field equations and
the connection associated to the Ricci tensor, which, due to the coupling
between the trace of the energy-momentum tensor and the geometry, is also a
function of the $T$. The metric that defines the new independent connection
is conformally related to the initial spacetime metric, with the conformal
factor given by a function of the trace of the energy-momentum tensor, and
of the Ricci scalar. After the conformal factor is obtained, the
gravitational field equations can be written down easily in both metrics.
Similarly to the case of the metric $f(R,T)$ gravity, after taking the
divergence of the gravitational field equations we obtain the important
result that the energy-momentum tensor of the matter is not conserved.
Similarly to the metric case, the motion of the particles is not geodesic,
and due to the matter-geometry coupling, an extra force arises. However,
this force has the same expression as in the metric case, and therefore no
new physics is expected to arise during the motion of massive test particles
in the Palatini formulation of the $f(R,T)$ gravity. As the next step in our
analysis we investigate in detail the cosmological implications of the
Palatini formulation of the $f(R,T)$ gravity theory. We obtain the
generalized Friedmann equations, which explicitly contain the extra terms
generated by the coupling between the trace of the energy-momentum tensor
and geometry. The general properties of the cosmological evolution are
obtained, including the behavior of the deceleration parameter, of the
effective energy density and pressure, and of the parameter of the equation
of state of the dark energy. Cosmological models with Lagrangians of the
type $L=R-\alpha ^2/R+g(T)$ and $L=R+\alpha ^2R^2+g(T)$ are considered in
detail, and it is shown that these models lead to evolution equations whose
solutions tend to a de Sitter type Universe at late times.

The present paper is organized as follows. After a brief review of the
metric formalism, the field equations of the $f(R,T)$ gravity theory are
obtained by using the Palatini formalism of gravitational theories in
Section~\ref{sect1}. The energy and momentum balance equations are obtained,
after taking the divergence of the energy-momentum tensor, in Section~\ref%
{sect2}. The thermodynamical interpretation of the theory is also briefly discussed. The cosmological implications of the Palatini $f(R,T)$ theory is
investigated in Section~\ref{sect3}. We discuss and conclude our results in
Section~\ref{sect4}. The details of the derivation of the field equations in
the metric formalism are given in Appendix~\ref{app1}, while the divergence
of the matter energy-momentum tensor is derived in Appendix~\ref{app2}. The
explicit computations of the various geometric quantities for the
Friedmann-Robertson-Walker geometry are presented in Appendix~\ref{app3}.

\section{Palatini formulation of $f(R,T)$ gravity}\label{sect1}

In the present Section, after a brief review of the metric formulation of
the $f(R,T)$ gravity theory, we derive the field equations of the theory by
using the Palatini formalism.

\subsection{The metric formalism}

The $f(R,T)$ gravity theory is described by the action \cite{Harkof}
\footnote{%
Throughout this article we use the natural system of units with $c = G = 1$.
For the metric tensor we adopt the signature convention $(-,+,+,+)$.}
\begin{equation}
S = \int \left[ \frac{\sqrt{-g}}{16 \pi} f(R, T) + \sqrt{-g}L_m \right] %
\thickspace \mathrm{d}^4 x,  \label{eq:fRTLm-act}
\end{equation}
where $g \equiv \mathrm{det} \left( g_{\mu\nu} \right)$, $f$ is an arbitrary
function of the Ricci scalar $R = R(g)$ and of the trace $T=g^{\mu
\nu}T_{\mu \nu}$ of the matter energy-momentum tensor $T_{\mu \nu}$; the
matter Lagrangian $L_m$ is assumed to be independent of $\partial_\lambda
g^{\mu\nu}$. $T_{\mu\nu}$ is generally obtained as \cite{L-CF}
\begin{eqnarray}
T_{\mu\nu} &\equiv & \frac{-2}{\sqrt{-g}} \left\{ \frac{\partial \left(
\sqrt{-g} L_m \right)}{\partial g^{\mu\nu}} + \partial_\lambda \left[ \frac{%
\partial \left( \sqrt{-g} L_m \right)}{\partial \left( \partial_\lambda
g^{\mu\nu} \right)} \right] \right\}  \notag \\
&=& \frac{-2}{\sqrt{-g}} \frac{\partial \left( \sqrt{-g} L_m \right)}{%
\partial g^{\mu\nu}} = - 2 \frac{\partial L_m}{\partial g^{\mu\nu}} +
g_{\mu\nu} L_m.  \label{eq:def-Tmunu}
\end{eqnarray}
To describe the variation of the energy-momentum tensor with respect to the
metric, we also introduce the tensor $\Theta_{\mu\nu}$, defined as
\begin{equation}
\Theta_{\mu\nu} \equiv g^{\alpha\beta} \frac{\delta T_{\alpha\beta}}{\delta
g^{\mu\nu}}.
\end{equation}

For a perfect fluid characterized by its energy density $\rho$ and isotropic
pressure $P$ only, the energy-momentum tensor is given by
\begin{equation}  \label{eq:perflu-Tmunu}
T_{\thickspace \nu}^\mu = (\rho + P) u^\mu u_\nu + P \delta_{\thickspace %
\nu}^\mu,
\end{equation}
where the four-velocity $u^{\mu}$ satisfies the normalization condition $%
u^\mu u_\mu = -1$. In the comoving frame its components are $u^\mu = (-1,
0,0,0)$, and in this frame the components of the energy-momentum tensor
become $T_{\thickspace \nu}^\mu=\left(-\rho,P,P,P\right)$.

The components of the affine connection are defined to be
\begin{equation}
\Gamma_{\thickspace \mu\nu}^{\rho} (g) = \frac{g^{\rho \sigma}}{2} \left(
\partial_\nu g_{\sigma\mu} + \partial_\mu g_{\sigma\nu} - \partial_\sigma
g_{\mu\nu} \right),  \label{eq:def-cons}
\end{equation}
which are currently regarded as functions of the metric. In the following we
assume that the connection is symmetric, that is, $\Gamma_{\thickspace %
\mu\nu}^{\rho} = \Gamma_{\thickspace \nu\mu}^{\rho}$. Varying Eq.~%
\eqref{eq:fRTLm-act} with respect to $g^{\mu\nu}$, we obtain
\begin{eqnarray}
&& \delta S = \int \frac{\sqrt{-g}}{16 \pi} \bigl[ f_R \delta R(g) + f_T
\delta T - \frac{g_{\mu\nu}}{2} f \delta g^{\mu\nu}  \notag \\
&& - 8 \pi T_{\mu\nu} \delta g^{\mu\nu} \bigr] \thickspace \mathrm{d}^4 x,
\end{eqnarray}
where $f_R \equiv \frac{\partial f}{\partial R}$, $f_T \equiv \frac{\partial
f}{\partial T}$ and
\begin{equation}
\delta R(g) = R_{\mu\nu} (g) \delta g^{\mu\nu} + g^{\mu\nu} \delta
R_{\mu\nu} (g).  \label{eq:deltaR=deltarmunugmunu}
\end{equation}
From the condition $\delta S = 0$ we obtain the field equations of the $%
f(R(g),T)$ gravity theory as (for the computational details see Appendix~\ref%
{app1}),
\begin{eqnarray}
&& \left[ R_{\mu\nu} (g) + g_{\mu\nu} \Box - \nabla_\mu \nabla_\nu \right]
f_R + \left( T_{\mu\nu} + \Theta_{\mu\nu} \right) f_T - \frac{g_{\mu\nu}}{2}
f  \notag \\
&& = 8 \pi T_{\mu\nu}.  \label{eq:feqm}
\end{eqnarray}

The field equations~\eqref{eq:feqm} can be rewritten with the help of the
Einstein tensor $G_{\mu\nu}(g) = R_{\mu\nu}(g) - g_{\mu\nu}R(g)/2$ as
\begin{eqnarray}
&& G_{\thickspace\nu}^\mu (g) = \frac{1}{f_{R}} \biggl\{ 8 \pi T_{\thickspace%
\nu}^\mu - \left( T_{\thickspace\nu}^\mu + \Theta_{\thickspace\nu}^\mu
\right) f_T + \frac{\delta_{\thickspace\nu}^\mu}{2} \left[ f - R (g) f_{R} %
\right]  \notag \\
&& + \left( \nabla^\mu \nabla_\nu - \delta_{\thickspace \nu}^\mu \Box
\right) f_{R} \biggr\}.  \label{eq:feqm1}
\end{eqnarray}

From Eq.~\eqref{eq:feqm1} it follows that the matter energy-momentum tensor
is not conserved, and its divergence is given by \cite{fRT5} (for the
computational details see Appendix~\ref{app2}),
\begin{eqnarray}  \label{eq:covder-Tmunum}
\nabla_{\mu} T_{\thickspace \nu}^\mu &=& \frac{f_{T}}{8\pi - f_{T}} \bigg[ %
\left( T_{\thickspace \nu}^\mu + \Theta_{\thickspace \nu}^\mu \right)
\nabla_{\mu} \ln \left| f_{T} \right| + \nabla_{\mu} \Theta_{\thickspace %
\nu}^\mu  \notag \\
&& - \frac{1}{2} \nabla_{\nu}T \bigg] \equiv Q_{\nu}.
\end{eqnarray}

\subsection{Palatini formulation of $f(R,T)$ gravity}

\subsubsection{Field equations from metric variation}

An alternative formulation of gravitational theories can be obtained within
the Palatini formalism, which \textit{consists in taking separately in the
gravitational action two independent variations, with respect to the metric
and the connection}, respectively. The action is \textit{formally identical
to the metric one}, but the \textit{Riemann tensor and the Ricci tensor are
constructed with the independent symmetric connection} $\tilde{\Gamma}$.
Hence in the Palatini formulation the gravitational action of the $f(R,T)$
gravity is given by
\begin{equation}
S = \int \left[ \frac{\sqrt{-g}}{16 \pi} f\left(R\left(g,\tilde{\Gamma}
\right), T\right) + \sqrt{-g}L_m\left(g,\psi\right) \right] \thickspace
\mathrm{d}^4 x.  \label{act_Pal}
\end{equation}
In Eq.~\eqref{act_Pal} the Ricci scalar is defined as
\begin{equation}  \label{riccsp}
R\left(g,\tilde{\Gamma}\right)=g^{\mu \nu} \tilde{R}_{\mu \nu}\left(\tilde{%
\Gamma}\right),
\end{equation}
with the Ricci tensor $\tilde{R}_{\mu \nu}\left(\tilde{\Gamma}\right)$
expressed only in terms of the Palatini connection $\tilde{\Gamma}$, with
the connection coefficients $\tilde{\Gamma}_{\mu \nu}^{\lambda}$ \textit{%
determined self-consistently through the independent variation of the
gravitational field action} Eq.~\eqref{act_Pal}, and \textit{not constructed
directly from the metric by using the usual Levi-Civita definition}. In the
following we define $\tilde{R}_{\mu \nu}\left(\tilde{\Gamma}\right)$ with
the help of the yet undetermined Palatini connection as
\begin{equation}  \label{ricctenp}
\tilde{R}_{\mu \nu}\left(\tilde{\Gamma}\right) =\partial _{\lambda }\tilde{%
\Gamma }_{\mu \nu }^{\lambda }-\partial _{\nu }\tilde{\Gamma }_{\mu \lambda
}^{\lambda }+\tilde{\Gamma }_{\mu \nu }^{\lambda }\tilde{\Gamma } _{\lambda
\alpha }^{\alpha }-\tilde{\Gamma }_{\mu \lambda }^{\alpha }\tilde{\Gamma }%
_{\nu \alpha }^{\lambda }.
\end{equation}
The matter Lagrangian $L_m\left(g,\psi\right)$ is assumed to be a function
of the metric tensor $g$ and of the physical fields $\psi$ only.

We vary now the gravitational action (\ref{act_Pal}) with respect to the
metric tensor $g^{\mu\nu}$, \textit{under the assumption} $\delta \tilde{R}%
_{\mu \nu}\left(\tilde{\Gamma}\right) = 0$, that is, \textit{by keeping the
connection constant}. As a result we immediately obtain the field equations
\begin{equation}  \label{eq:feqp}
\tilde{R}_{\mu \nu}\left(\tilde{\Gamma}\right) f_R = 8 \pi T_{\mu \nu} -
\left( T_{\mu\nu} + \Theta _{\mu\nu} \right)f_T + \frac{g_{\mu\nu}}{2} f.
\end{equation}

By contracting the above equation with $g^{\mu \nu}$ we obtain for the Ricci
scalar~\eqref{riccsp} the expression
\begin{equation}  \label{riccsp1}
R\left(g,\tilde{\Gamma}\right)f_R = 8\pi T - (T + \Theta) f_T + 2f,
\end{equation}
where $\Theta =\Theta _{\thickspace \mu}^{\mu}$. In term of the Einstein
tensor
\begin{equation}
G_{\mu \nu} \left(g,\tilde{\Gamma}\right) = \tilde{R}_{\mu \nu}\left(\tilde{%
\Gamma}\right) - \frac{1}{2}g_{\mu \nu} R \left(g,\tilde{\Gamma}\right),
\end{equation}
the Palatini field equations can be written as
\begin{eqnarray}  \label{etenp}
&& G_{\mu \nu} \left(g,\tilde{\Gamma}\right) = \frac{1}{f_R}\Bigg\{ 8\pi
\left( T_{\mu \nu} - \frac{g_{\mu\nu}}{2} T\right)  \notag \\
&& - \left[ \left( T_{\mu \nu } + \Theta_{\mu\nu} \right) - \frac{g_{\mu\nu}%
}{2} \left( T + \Theta \right) \right] f_{T} - \frac{g_{\mu\nu}}{2} f \Bigg\}%
.
\end{eqnarray}

\subsubsection{The Palatini connection}

We vary now the gravitational action~\eqref{act_Pal} with respect to the
connection $\tilde{\Gamma}$, \textit{by keeping the metric constant}, so
that
\begin{equation}
\delta R \left( g, \tilde{\Gamma} \right) = g^{\mu\nu} \delta \tilde{R}_{\mu
\nu}\left(\tilde{\Gamma}\right).
\end{equation}
According to Palatini identity \cite{H-Pl}, we have
\begin{equation}
\delta \tilde{R}_{\mu \nu}\left(\tilde{\Gamma}\right) =\tilde{\nabla}
_{\lambda }\left( \delta \tilde{\Gamma }_{\mu \nu }^{\lambda }\right) -%
\tilde{\nabla }_{\nu }\left( \delta \tilde{\Gamma }_{\mu \lambda}^{\lambda
}\right),
\end{equation}
where $\tilde{\nabla}_{\lambda}$ is the covariant derivative associated with
$\tilde{\Gamma}$. Hence the variation of the action~\eqref{act_Pal} with
respect to $\tilde{\Gamma }$ leads to
\begin{equation}
\delta S =\int \frac{\sqrt{-g}}{16 \pi} A^{\mu \nu }\left[ \tilde{\nabla }
_{\lambda }\left( \delta \tilde{\Gamma }_{\mu \nu }^{\lambda }\right) -%
\tilde{\nabla }_{\nu }\left( \delta \tilde{\Gamma }_{\mu \lambda }^{\lambda
}\right) \right] \thickspace \mathrm{d}^{4}x,
\end{equation}
where we have denoted
\begin{equation}
A^{\mu \nu }=f_R g^{\mu \nu }.
\end{equation}

We integrate now by parts to obtain
\begin{eqnarray}  \label{19}
&& 16 \pi \delta S = \int \tilde{\nabla }_{\lambda }\left[ \sqrt{-g}\left(
A^{\mu \nu }\delta \tilde{\Gamma }_{\mu \nu }^{\lambda
}-A^{\mu\lambda}\delta \tilde{\Gamma }_{\mu \alpha }^{\alpha }\right) \right]
\thickspace \mathrm{d}^{4}x  \notag \\
&& - \int \tilde{\nabla }_{\lambda }\left[ \sqrt{-g}\left( A^{\mu \nu
}\delta _{\alpha }^{\lambda }-A^{\mu\lambda }\delta _{\alpha }^{\nu }\right) %
\right] \delta \tilde{\Gamma }_{\mu \nu }^{\alpha } \thickspace \mathrm{d}%
^{4}x.
\end{eqnarray}

The first term in Eq.~(\ref{19}) is a total derivative, and thus after
transforming it into a surface integral it vanishes. Therefore the variation
of the action with respect to the connection $\tilde{\Gamma}$ becomes
\begin{equation}
\tilde{\nabla }_{\lambda }\left[ \sqrt{-g}\left( A^{\mu \nu }\delta _{\alpha
}^{\lambda }-A^{\mu\lambda }\delta _{\alpha }^{\nu }\right) \right] =0.
\label{con}
\end{equation}
Eq.~(\ref{con}) can be significantly simplified by taking into account that
for $\alpha =\lambda $ the equation is identically zero. Hence for the case $%
\lambda \neq \alpha $, we find
\begin{equation}  \label{con1}
\tilde{\nabla}_{\lambda} \left( \sqrt{-g} f_R g^{\mu\nu} \right) = 0.
\end{equation}
Eq.~\eqref{con1} shows that the connection $\tilde{\Gamma}$ \textit{is
compatible with a conformal metric} $\tilde{g}_{\mu \nu}$, conformally
related to the initial metric $g_{\mu \nu}$ by means of the relations
\begin{equation}  \label{eq:newmetric}
\tilde{g}_{\mu\nu} \equiv f_Rg_{\mu\nu} = Fg_{\mu \nu},
\end{equation}
with the conformal factor $F$ defined as
\begin{equation}
F \equiv f_R \left( R\left(g,\tilde{\Gamma}\right), T \right).
\label{eq:confac}
\end{equation}
Moreover, we have
\begin{equation}
\sqrt{-\tilde{g}}\tilde{g}^{\mu \nu }=\sqrt{-g}f_R g^{\mu \nu } = \sqrt{-
F^4 g} \left( F^{-1} g^{\mu\nu} \right),
\end{equation}
where $\tilde{g} \equiv \det \left( \tilde{g}_{\mu \nu} \right)$. Thus Eq.~%
\eqref{con1} gives \textit{the geometric interpretation of the Palatini
connection} $\tilde{\Gamma }$ as the Levi-Civita connection corresponding to
the metric $\tilde{g}_{\mu \nu }$, conformally related to $g_{\mu \nu}$,
\begin{equation}
\tilde{\Gamma }_{\mu \nu }^{\lambda }=\frac{1}{2}\tilde{g}^{\lambda \rho
}\left( \partial _{\nu }\tilde{g}_{\rho\mu }+\partial _{\mu }\tilde{g}%
_{\rho\nu }-\partial _{\rho }\tilde{g}_{\mu \nu }\right).
\label{eq:Pal-conn}
\end{equation}

\subsubsection{Conformal geometry and $g$ frame field equations}

Since the metrics $g_{\mu \nu}$ and $\tilde{g}_{\mu \nu}$ are conformally
related, the connection $\tilde{\Gamma}$ can be expressed in terms of the
Levi-Civita connection $\Gamma$, whose components are given in Eq.~%
\eqref{eq:def-cons}, as
\begin{equation}
\tilde{\Gamma}_{\mu \nu }^{\lambda } =\Gamma _{\mu \nu }^{\lambda } + \frac{1%
}{2F} \left( \delta _{\thickspace \mu}^{\lambda }\partial _{\nu } + \delta_{%
\thickspace \nu}^{\lambda }\partial_{\mu } - g_{\mu \nu }\partial^{\lambda }
\right) F.  \label{eq:Pal-conn0}
\end{equation}
In terms of the tensor $R_{\mu \nu}(g)$, constructed from the metric by
using the Levi-Civita connection~\eqref{eq:def-cons}, the Ricci tensor $%
\tilde{R}_{\mu \nu}$ in the conformally transformed metric is given by \cite%
{Koi, X-GRG}
\begin{equation}  \label{rictenc}
\tilde{R}_{\mu \nu} = R_{\mu \nu}(g) + \frac{1}{F} \left[ \frac{3}{2F}
\nabla_\mu F \nabla_{\nu}F - \left( \nabla_{\mu} \nabla_{\nu} + \frac{g_{\mu
\nu}} {2} \Box \right) F \right].
\end{equation}
Since $\tilde{R} \equiv \tilde{g}^{\mu\nu} \tilde{R}_{\mu\nu} = F^{-1} R
\left(g,\tilde{\Gamma}\right)$ and $\tilde{G}_{\mu\nu} \equiv \tilde{R}%
_{\mu\nu} - \tilde{g}^{\mu\nu} \tilde{R} /2 = G_{\mu\nu} \left(g,\tilde{%
\Gamma}\right)$, the Ricci scalar~\eqref{riccsp} and the Einstein tensor in
Eq.~\eqref{etenp} can be obtained in the conformally related frames as \cite%
{Koi,X-GRG}
\begin{equation}  \label{ricscalc}
R\left(g,\tilde{\Gamma}\right) = F \tilde{R} = R (g) + \frac{3}{F} \left[
\frac{1}{2F} \left(\nabla F\right)^2 - \Box F \right]
\end{equation}
and
\begin{eqnarray}  \label{ein}
&&G_{\mu \nu} \left(g,\tilde{\Gamma}\right) = \tilde{G}_{\mu \nu } =G_{\mu
\nu }(g) + \frac{1}{F} \bigg\{ \left( g_{\mu \nu}\Box - \nabla _{\mu }\nabla
_{\nu } \right) F  \notag \\
&& + \frac{3}{2F} \Big[ \nabla_{\mu} F \; \nabla_{\nu}F - \frac{g_{\mu \nu}}{%
2} \left(\nabla F\right)^2 \Big] \bigg\}
\end{eqnarray}
respectively, \textit{with all the covariant derivatives and algebraic
operations performed with the help of the metric $g_{\mu \nu }$}.

By using the expression of the Einstein tensor as given by Eq.~\eqref{ein}
in Eq.~\eqref{etenp}, we obtain finally the $f(R,T)$ field equations of the
Palatini formulation \textit{expressed solely in the $g$ frame} as
\begin{eqnarray}
&&G_{\mu \nu }(g) + \frac{1}{F} \bigg\{ \left( g_{\mu \nu}\Box - \nabla
_{\mu }\nabla _{\nu } \right) F + \frac{3}{2F} \Big[ \nabla_{\mu} F \;
\nabla_{\nu}F  \notag \\
&& - \frac{g_{\mu \nu}}{2} \left(\nabla F\right)^2 \Big] \bigg\} = \frac{1}{F%
}\Bigg\{ 8\pi \left( T_{\mu \nu} - \frac{g_{\mu\nu}}{2} T\right)  \notag \\
&& - \left[ \left( T_{\mu \nu } + \Theta_{\mu\nu} \right) - \frac{g_{\mu\nu}%
}{2} \left( T + \Theta \right) \right] f_{T} - \frac{g_{\mu\nu}}{2} f \Bigg\}%
.  \label{eq:feqp1}
\end{eqnarray}

The Ricci scalar can be obtained as
\begin{eqnarray}  \label{feqp-tr}
&&R\left( g\right) + \frac{3}{F} \left[ \frac{1}{2F} \left(\nabla F\right)^2
- \Box F \right] =  \notag \\
&&\frac{1}{F}\left[8 \pi T - \left( T + \Theta \right) f_{T} + 2f \right] .
\end{eqnarray}

\subsubsection{Field equations in the $\tilde{g}$ frame}

In the previous Subsection we have obtained the gravitational field
equations in the Palatini formulation of $f(R,T)$ gravity as expressed in
terms of the metric tensor $g_{\mu \nu}$. This approach usually involves
higher order differential equations for the physical and geometrical
quantities. An alternative approach, which keeps the order of the
differential equations of the model not higher than two would be to solve
first the gravitational field equations in the conformal metric $\tilde{g}%
_{\mu \nu}$, and to recover the metric $g_{\mu \nu}$ with the use of the
conformal transformation (\ref{eq:newmetric}). To obtain the field equations
in the conformal frame we follow the procedure introduced in \cite{Pal1}.

First we multiply the Palatini field equation~\eqref{eq:feqp} with $\tilde{g}%
^{\lambda \mu}={f_R}^{-1} g^{\lambda \mu}$, thus obtaining
\begin{equation}  \label{37}
\tilde{R}_{\thickspace \nu}^{\lambda} = \frac{1}{F^2}\left[8\pi T_{%
\thickspace \nu}^{\lambda} - \left(T_{\thickspace \nu}^{\lambda} + \Theta_{%
\thickspace \nu}^{\lambda} \right)f_T+\frac{\delta_{\thickspace %
\nu}^{\lambda}}{2} f \right],
\end{equation}
where $\tilde{R}_{\thickspace \nu}^{\lambda}$ is now a function of the
metric $\tilde{g}$ only. As for the energy-momentum tensor, we have,
respectively,
\begin{equation}
T_{\mu\nu} \rightarrow \tilde{T}_{\mu\nu} = \frac{-2}{\sqrt{- \tilde{g}}}
\frac{\partial \left( \sqrt{- \tilde{g}} L_m \right)}{\partial \tilde{g}%
^{\mu\nu}} = - 2 \frac{\partial L_m}{\partial \tilde{g}^{\mu\nu}} + \tilde{g}%
_{\mu\nu} L_m
\end{equation}
and
\begin{equation}
T_{\thickspace\nu}^\mu \rightarrow \tilde{T}_{\thickspace\nu}^\mu = - 2
\tilde{g}^{\mu\lambda} \frac{\partial L_m}{\partial \tilde{g}^{\lambda\nu}}
+ \delta_{\thickspace\nu}^\mu L_m = T_{\thickspace\nu}^\mu.
\label{eq:invTmunu}
\end{equation}
That is, the mixed components $T_{\thickspace\nu}^\mu$ of the
energy-momentum tensor are conformally invariant. Additionally, we have
\begin{equation}
\Theta_{\thickspace\nu}^\mu \rightarrow \tilde{\Theta}_{\thickspace\nu}^\mu
= \Theta_{\thickspace\nu}^\mu, \quad T \rightarrow \tilde{T} = T, \quad
\Theta \rightarrow \tilde{\Theta} = \Theta.  \label{eq:invT&theta}
\end{equation}

By contracting Eq.~(\ref{37}) by taking $\nu =\lambda$ we obtain the
expression of the Ricci scalar in the conformal frame as
\begin{equation}  \label{riccsc}
\tilde{R} = \frac{1}{F^2}\left[8\pi T-\left(T+\Theta\right)f_T+2f\right].
\end{equation}
Therefore in the conformal frame the full set of the Einstein equations can
be written with $f = f\left(F \tilde{R} = R\left(g,\tilde{\Gamma}\right),
T\right)$ as
\begin{eqnarray}
\hspace{-0.6cm}\tilde{G}_{\thickspace \nu}^{\mu} &=& \tilde{R}_{\thickspace %
\nu}^{\mu} - \frac{\delta_{\thickspace \nu}^{\mu}}{2} \tilde{R} = \frac{1}{%
F^2}\Bigg\{ 8\pi \left( T_{\thickspace \nu}^{\mu} - \frac{\delta_{%
\thickspace \nu}^{\mu}}{2} T\right)  \notag \\
\hspace{-0.6cm}&& - \left[ \left( T_{\thickspace \nu}^{\mu} + \Theta_{%
\thickspace \nu}^{\mu} \right) - \frac{\delta_{\thickspace \nu}^{\mu}}{2}
\left( T + \Theta \right) \right] f_{T} - \frac{\delta_{\thickspace %
\nu}^{\mu}}{2} f \Bigg\}.  \label{eq:feqc}
\end{eqnarray}

These equations determine the conformal metric $\tilde{g}$ as a function of
the thermodynamic parameters that enter in the definition of the matter
energy-momentum tensor.

\subsection{The Newtonian limit}

To investigate the  Palatini $f(R,T)$ gravity under the weak field, slow motion and static approximation, namely the Newtonian limit, we assume the metric to be a Minkowski metric plus a perturbation, given by
\begin{equation} \label{eq:newtonianmetric}
g_{\mu \nu} = \eta_{\mu \nu} + \gamma_{\mu \nu},
\quad \gamma_{\mu \nu} \ll 1.
\end{equation}
Hence $g^{\mu\nu} = \eta^{\mu \nu} - \gamma^{\mu \nu}$ and the relation $g^{\alpha \lambda}g_{\lambda \beta}=\delta^{\alpha}_{\; \beta}$ still holds. In this context, we also assume the conformal metric~\eqref{eq:newmetric} to be nearly flat, so that
\begin{equation}
\label{eq:conformalnewtonianmetric}
\tilde{g}_{\mu \nu} = F g_{\mu \nu} = \eta_{\mu \nu} + \tilde{\gamma}_{\mu \nu},
\end{equation}
where $\tilde{\gamma}_{\mu \nu} \ll 1$ is of the same order as $\gamma_{\mu \nu}$. Hence $F \approx 1$. From Eqs.~\eqref{eq:newtonianmetric} and \eqref{eq:conformalnewtonianmetric} we obtain
\begin{equation} \label{eq:orderofw}
(F-1)\eta_{\mu\nu}
= \tilde{\gamma}_{\mu \nu} - F \gamma_{\mu\nu}.
\end{equation}
If we take $F = e^{2W}$ and expand it to $F=1+2W$, then the above equation yields that $W =\left(\tilde{\gamma} - \gamma\right)/2 \left( 4 + \gamma \right)$, where $\tilde{\gamma} \equiv \eta^{\mu\nu} \tilde{\gamma}_{\mu \nu}$ and $\gamma \equiv \eta^{\mu \nu} \gamma_{\mu \nu}$. Thus $W \sim O(\gamma) \sim O(\gamma_{\mu\nu})$.

Let us now consider the Palatini field equations~\eqref{eq:feqp} under the Newtonian limit. First we obtain the $g$-frame Ricci tensor as
\begin{equation}  \label{eq:newtonianricci}
R_{\mu\nu}(g)=\frac{1}{2} \left(\partial^{\lambda}\partial_{\mu} \gamma_{\nu
\lambda} + \partial^{\lambda}\partial_{\nu} \gamma_{\mu
\lambda} - \partial^2 \gamma_{\mu \nu} -
\partial_{\mu\nu} \gamma \right).
\end{equation}
Omitting all higher-order terms with respect to $O(\gamma_{\mu \nu})$ and taking into account the gauge
\be
\partial^\mu \left( \gamma_{\mu \nu} - \frac{1}{2}\eta_{\mu \nu}\gamma \right)=0,
\ee
then some algebra gives the expression of the Ricci tensor~\eqref{rictenc} as
\begin{equation}  \label{eq:conformalnewtonianricci}
\tilde{R}_{\mu \nu}
= - \frac{1}{2} \partial^2 \left( \gamma_{\mu \nu} +2\eta_{\mu \nu} w \right) - 2 \partial_{\mu\nu} w .
\end{equation}
With the use of the above equations, and preserving only the first-order terms, the field equations~\eqref{eq:feqp} become
\begin{equation}  \label{eq:newtonianfeq}
-\frac{1}{2} \partial^2 \tilde{\gamma}_{\mu \nu} - 2\partial_{\mu\nu} w -\frac{\eta_{\mu \nu}}{2}f = 8 \pi T_{\mu
\nu} - \left( T_{\mu\nu} + \Theta _{\mu\nu} \right)f_T .
\end{equation}
For perfect fluids, $T^{\mu}_{\; \nu}=\mathrm{diag}(-\rho,P,P,P)$ and $\Theta_{\thickspace \nu}^\mu = \delta_{\thickspace \nu}^\mu P - 2 T_{\thickspace \nu}^\mu$ (see Eq.~\eqref{eq:Thetamunu2}); besides, under the Newtonian limit, $P \to 0$ and $\partial_0 \to 0$. Hence we can obtain immediately the generalized Poisson equation in the Palatini formulation of $f(R,T)$ gravity as
\begin{equation}  \label{eq:possion}
-\frac{1}{2} \vec{\nabla}^2 \tilde{\gamma}_{0 0} = \left( 8 \pi + f_T \right) \rho  -\frac{f}{2}.
\end{equation}
The same result can also be found in \cite{Pal1}.

\subsection{Violation of the equivalence principle}

An interesting feature of the modified gravities, including their Palatini extensions, is the violation of the equivalence principle. In the Palatini $f(R)$ gravity this problem was discussed in \cite{Olmo_new}. In the following we will generalize some results from the case of the $f(R)$ theory to the Palatini $f(R,T)$ gravity. In the conformal frame the field equations of the Palatini $f(R,T)$ gravity are given by Eqs.~(\ref{eq:feqc}), where $f = f\left(F \tilde{R},T\right) =f\left( R\left(g,\tilde{\Gamma}\right),T\right)$.

In the weak field limit we can represent the gravitational Lagrangian as
 \be
f\left( R\left(g,\tilde{\Gamma}\right),T\right) = R(g,\tilde{\Gamma}) + \epsilon K\Big(R\left(g,\tilde{\Gamma}\right)\Big) + \epsilon' g(T).
\ee
where $\epsilon$, $\epsilon'$ are constants, $K\Big(R(g,\tilde{\Gamma})\Big)$ is an arbitrary function of the argument $R(g,\tilde{\Gamma})$, while $g$ is an arbitrary function of the trace of the matter energy-momentum tensor. In the limit of small $\epsilon, \epsilon'$, with  $\epsilon, \epsilon' \rightarrow 0$, it follows that $f-R(g,\tilde{\Gamma})F \approx 0$. In addition, by neglecting the matter energy-momentum tensor $T_{\;\nu}^\mu$ for weak sources, it follows that Eq.~\eqref{eq:feqc} becomes $\tilde{G}_{\;\nu}^\mu \approx 0$, which leads to $\tilde{g}_{\mu\nu} \approx \eta_{\mu\nu}$, or, equivalently,
\begin{equation}
  g_{\mu\nu} \approx F^{-1} \eta_{\mu\nu} \approx \left(1-\epsilon \frac{\partial K}{\partial R}\right) \eta_{\mu\nu}.
\end{equation}
The above equation tells us that, similarly to the metric and Palatini formulation of $f(R)$ gravity, \cite{Olmo_new}, \textit{in the Palatini formulation of $f(R,T)$ gravity it is impossible to recover the flat Minkowski metric even in local frames with external gravitational fields screened}. This result violates the basic postulate of general relativity according to which \textit{in locally freely falling frames the non-gravitational laws of physics are those of special relativity} \cite{Olmo_new}. Since this postulate assumes that the Einstein Equivalence Principle holds, it follows that
that similarly to the Palatini $f(R)$ theory, in the Palatini formulation of $f(R,T)$ gravity the equivalence principle does not hold exactly. The deviation of the current metric with respect to the Minkowski metric is $\eta_{\mu\nu} - g_{\mu\nu}\approx \epsilon \left(\partial K/\partial R\right) \eta_{\mu\nu}$.

In order to give a quantitative estimate for the deviation of the $f(R,T)$ metric from the Minkowski metric we will consider the cosmological case, to be discuss in detail in Section~\ref{sect3}. We assume that the gravitational action takes  the form $f(R(g,\tilde{\Gamma}),T) = R(g,\tilde{\Gamma}) + \frac{\alpha}{16 \pi} R^2 (g,\tilde{\Gamma}) + 8 \pi \beta T$, where $\alpha,\beta\rightarrow0$, as an example. From Eq.~(\ref{Fm1}) to be derived in Section~\ref{sect3}, it follows that $F = 1 + \beta_0 \alpha \rho =  1 + \beta_0 \alpha \rho_0 a^{- \beta_1}$, where $\rho _0$ is the present day matter density, $a$ is the scale factor, $\beta_0 = 1 - 3w + \beta (3 - 5w)$, $\beta_1 = 3(1+\beta)(1+w)/[1+\beta(3-w)/2]$, and $w$ is the parameter of the matter equation of state. By estimating  all quantities at the present time $t=t_0$, then when $\alpha,\beta \rightarrow0$, and $w=0$, $a\left(t_0\right)=1$,  the deviation $\epsilon \left(\partial K/\partial R\right)$ from the Minkowski metric can be obtained as
\bea
1 - F^{-1} &=& 1 -[1 + \left.\beta_{0}\right|_{t=t_0} \alpha \rho_0 a_0^{- \left.\beta_{1}\right|_{t=t_0}}]^{-1}
\approx \nonumber\\
&&\left(1+3\beta\right) \alpha \rho_0
\approx \alpha \rho _0\approx \alpha \frac{3H_0^2}{8\pi },
\eea
where $H_0$ is the present day value of the Hubble function, and we have assumed for the present-day density of the Universe the critical value.

Recently the first results obtained by the MICROSCOPE satellite, whose  aims are to constrain the Weak Equivalence Principle in the outer space by determining the E\"{o}tv\"{o}s parameter $\eta $, have been published \cite{Micr1}. The E\"{o}tv\"{o}s parameter is defined as the normalized difference of accelerations between two bodies $i$ and $j$, located in the same gravitational field. The MICROSCOPE determinations  give for $\eta $ the value $\eta = (-1 \pm 27) \times 10^{-15}$ at a 2-$\sigma$ confidence level \cite{Micr1}. These results allow to constrain possible sources of violation of the weak equivalence principle, like,  for example, the existence of  light or massive scalar fields with coupling to matter weaker than the gravitational
coupling \cite{Micr2}. For a massive scalar field of mass smaller than $10^{-12}$ eV,  the coupling is constrained as $\left|\alpha _C\right|<10^{-11}$, if the scalar field couples to the baryon number, and to $\left|\alpha _C\right|<10^{-12}$ if the scalar field couples to the difference between the baryon and the lepton numbers, respectively. We expect similar order of magnitude for the coupling between matter and geometry in both metric and Palatini formulations of $f(R,T)$ gravity.

\section{Energy and momentum balance equations}

\label{sect2}

An interesting and important consequence of modified gravity theories with
geometry-matter coupling is the non-conservation of the matter
energy-momentum tensor. This property of the theory has a number of far
reaching physical implications, and may represent the main link between the
interpretation of the $f(R,T)$ theory as an effective classical description
of the quantum theory of gravity. In the present section we obtain the
general expression of the divergence of the energy-momentum tensor in $%
f(R,T) $ gravity theory, and, by using it, we obtain the energy-momentum
balance equations, which describe the energy transfer processes from
geometry to matter, and the deviations from the geodesic motion,
respectively.

\subsection{The divergence of the matter energy-momentum tensor}

\label{sec:non-conserv-Tmunu}

We begin our analysis by calculating first the divergence of the Einstein
tensor in the Palatini frame. Since $\nabla^\mu G_{\mu\nu} (g) = 0$, then
Eqs.~\eqref{ein} yields
\begin{eqnarray}
&& \nabla^\mu \tilde{G}_{\mu\nu} = \nabla^\mu \left[ \tilde{G}_{\mu\nu} -
G_{\mu\nu} (g) \right]  \notag \\
&=& 2 \left( \nabla_\nu \Box - \Box \nabla_\nu \right) w + 2 \Box w
\nabla_\nu w + 2 \nabla^{\mu} w \nabla_\mu \nabla_\nu w  \notag \\
&& + \nabla_\nu \left( \nabla w \right)^2  \notag \\
&=& - 2 R_{\mu \nu} (g) \nabla^\mu w + 2 \Box w \nabla_\nu w + 4
\nabla^{\mu} w \nabla_\mu \nabla_\nu w,  \label{eq:nablaEtens}
\end{eqnarray}
where to simplify the calculation we have taken $F = e^{2w}$, and we
have used the mathematical identities \cite{Koi}
\begin{eqnarray}
&& \left( \nabla_\nu \Box - \Box \nabla_\nu \right) \phi = g^{\alpha\beta}
\left( \nabla_\nu \nabla_\alpha \nabla_\beta - \nabla_\alpha \nabla_\beta
\nabla_\nu \right) \phi  \notag \\
&=& g^{\alpha\beta} \left( \nabla_\nu \nabla_\alpha - \nabla_\alpha
\nabla_\nu \right) \nabla_\beta \phi = g^{\alpha \beta} R_{\thickspace \beta
\alpha \nu}^\mu \nabla_\mu \phi  \notag \\
&=& - R_{\mu \nu} \nabla^\mu \phi  \label{eq:com-box-nabla}
\end{eqnarray}
and
\begin{eqnarray}
\nabla_\nu \left( \nabla \phi \right)^2 &=& \nabla_\nu \left( \nabla^\mu
\phi \nabla_\mu \phi \right) = \nabla_\mu \phi \nabla_\nu \nabla^\mu \phi +
\notag \\
&&\nabla^\mu \phi \nabla_\nu \nabla_\mu \phi = 2 \nabla_\mu \phi \nabla_\nu
\nabla^\mu \phi  \notag \\
&=& 2 \nabla^\mu \phi \nabla_\nu \nabla_\mu \phi,
\end{eqnarray}
respectively, where $\phi \left( x^\mu \right)$ is an arbitrary scalar. We
have also used the relation $g^{\alpha \beta} R_{\mu \beta \alpha \nu} = -
g^{\alpha \beta} R_{\beta \mu \alpha \nu} = - R_{\mu \nu} $ in the last step
of Eq.~\eqref{eq:com-box-nabla}. Note that the above two identities are
valid for in both the metric and the Palatini formulations \cite{Koi}.

Substituting Eq.~\eqref{rictenc} into Eq.~\eqref{eq:nablaEtens}, we find
\begin{eqnarray}
\nabla^\mu \tilde{G}_{\mu\nu} &=& - 2 R_{\mu\nu} (g)\nabla^\mu w + 2 \Box w
\nabla_\nu w + 4 \nabla^{\mu} w \nabla_\mu \nabla_\nu w  \notag \\
&=& - 2 \tilde{R}_{\mu\nu} \nabla^\mu w = - \frac{\nabla^\mu F}{F} \tilde{R}%
_{\mu\nu}.  \label{eq:nablaEtens2}
\end{eqnarray}

The covariant divergence of the field equations~\eqref{etenp} yields (note
that $G_{\mu \nu} \left(g,\tilde{\Gamma}\right) = \tilde{G}_{\mu\nu}$ and $%
g_{\mu\nu}R \left(g,\tilde{\Gamma}\right) = \tilde{g}_{\mu\nu} \tilde{R}$),
\begin{eqnarray}  \label{eq:covder-Tmunup1a}
\hspace{-0.5cm}&& \nabla^\mu \left[ F G_{\mu \nu} \left(g,\tilde{\Gamma}%
\right) \right] + \frac{g_{\mu\nu}}{2} R \left(g,\tilde{\Gamma}\right)
\nabla^\mu F  \notag \\
\hspace{-0.2cm}&=& \left[ \tilde{G}_{\mu \nu} - \tilde{R}_{\mu \nu} + \frac{%
\tilde{g}_{\mu\nu}}{2} \tilde{R} \right] \nabla^\mu F = 0  \notag \\
\hspace{-0.2cm}&=& \nabla_\mu \left[ 8 \pi T_{\thickspace\nu}^\mu - \left(
T_{\thickspace \nu}^\mu + \Theta_{\thickspace\nu}^\mu \right) f_T \right] +
\notag \\
&&\frac{\delta_{\thickspace\nu}^\mu}{2} \left[ \nabla_\mu f - F \nabla_\mu R
\left(g,\tilde{\Gamma}\right) \right]  \notag \\
\hspace{-0.2cm}&=& \nabla_\mu \left[ 8 \pi T_{\thickspace\nu}^\mu - \left(
T_{\thickspace\nu}^\mu + \Theta_{\thickspace\nu}^\mu \right) f_T \right] +
\frac{f_T}{2} \nabla_\nu T,  \label{eq:covder-Tmunup}
\end{eqnarray}
where we have used Eqs.~\eqref{eq:nablaEtens2}.

One can check now by comparison with Eq.~\eqref{eq:covder-Tmunum} that the
above expression gives the same result as in the metric formulation, except
for the functional form of the function $f$,
\begin{eqnarray}  \label{eq:covder-Tmunup1}
\nabla_{\mu} T_{\thickspace \nu}^\mu &=& \frac{f_{T}}{8\pi - f_{T}} \bigg[ %
\left( T_{\thickspace \nu}^\mu + \Theta_{\thickspace \nu}^\mu \right)
\nabla_{\mu} \ln \left| f_{T} \right| + \nabla_{\mu} \Theta_{\thickspace %
\nu}^\mu  \notag \\
&& - \frac{1}{2} \nabla_{\nu}T \bigg], \quad f = f\left(R\left(g,\tilde{%
\Gamma}\right), T\right).
\end{eqnarray}

From its definition with the use of Eq.~\eqref{eq:def-Tmunu}, the tensor $%
\Theta_{\mu \nu}$ for perfect fluids can be obtained as
\begin{eqnarray}
\Theta_{\mu\nu} &=& g^{\alpha\beta} \left[ \frac{\delta g_{\alpha\beta}}{%
\delta g^{\mu\nu}} L_m + g_{\alpha\beta} \frac{\partial L_m}{\partial
g^{\mu\nu}} - 2 \frac{\partial^2 L_m}{\partial g^{\mu\nu} \partial
g^{\alpha\beta}} \right]  \notag \\
&=& g_{\mu\nu} L_m - 2 T_{\mu\nu} - 2 g^{\alpha\beta} \frac{\partial^2 L_m} {%
\partial g^{\mu\nu} \partial g^{\alpha\beta}} ,  \label{eq:Thetamunu}
\end{eqnarray}
where the relation
\begin{equation}
\frac{\delta g_{\alpha\beta}}{\delta g^{\mu\nu}} = - g_{\alpha\mu}
g_{\beta\nu},
\end{equation}
can be derived from the relations
\begin{eqnarray}
&&g^{\alpha\beta} \left( \delta g_{\alpha\beta} + g_{\alpha\mu} g_{\beta
\nu} \delta g^{\mu\nu} \right) = g^{\alpha\beta} \delta g_{\alpha\beta} +
g_{\mu\nu} \delta g^{\mu\nu} = 0  \notag \\
&&\Rightarrow \delta g_{\alpha\beta} = - g_{\alpha\mu} g_{\beta \nu} \delta
g^{\mu\nu}.
\end{eqnarray}

For perfect fluids we fix $L_m$ to be $P$ \cite{fRT5}, while $T_{\thickspace %
\nu}^\mu$ takes the form of \eqref{eq:perflu-Tmunu}, then
\begin{equation}  \label{eq:Thetamunu2}
\Theta_{\mu\nu} = g_{\mu\nu} P - 2 T_{\mu\nu},
\quad \Theta_{\thickspace \nu}^\mu = \delta_{\thickspace \nu}^\mu P - 2 T_{\thickspace \nu}^\mu;
\end{equation}
\begin{equation}  \label{eq:perflu-T&theta}
T = \delta_{\thickspace \mu}^\nu T_{\thickspace \nu}^\mu = - \rho + 3 P,
\quad \Theta \equiv \delta_{\thickspace \mu}^\nu \Theta_{\thickspace %
\nu}^\mu = 4 P - 2 T.
\end{equation}
Hence
\begin{equation}
T_{\mu\nu} + \Theta_{\mu\nu} = g_{\mu\nu} P - T_{\mu\nu},
\end{equation}
and
\begin{equation}
T + \Theta = 4 P - T = \rho + P.
\end{equation}
Multiplying Eq.~\eqref{eq:covder-Tmunup1} by $u^\nu$ \cite{Ha14} we obtain
the $f(R,T)$ perfect-fluid energy balance equation
\begin{eqnarray} \label{61}
&&\dot{\rho}+3(\rho + P) H = \frac{-f_T}{8\pi +f_T} \bigg[(\rho + P) u^{\mu}
\nabla_{\mu} \ln \left| f_T \right|  \notag \\
&& + u^{\mu} \nabla_{\mu} \left( \frac{\rho - P}{2} \right) \bigg], \quad f
= f \left( R\left(g,\tilde{\Gamma}\right), T \right),
\end{eqnarray}
where we have denoted $H = \left( \nabla_{\mu} u^{\mu} \right)/3$, and $\dot{%
} = \mathrm{d} / \mathrm{d} s = u^{\mu } \nabla_{\mu }$.

Multiplying \eqref{eq:covder-Tmunup1} by the projection operator $h_{%
\thickspace \lambda}^\nu$, defined as $h_{\thickspace \lambda}^\nu \equiv
\delta_{\thickspace \lambda}^\nu + u^\nu u_\lambda $ \cite{Ha14}, with the
properties $u_\nu h_{\thickspace \lambda}^\nu = 0$, $h_{\thickspace %
\lambda}^\nu \nabla_\mu u_\nu = \nabla_\mu u_\lambda$, and $h^{\nu \lambda}
\nabla_\nu = \left( g^{\nu\lambda} + u^\nu u^\lambda \right) \nabla_\nu =
\nabla^\lambda + u^\lambda u^\nu \nabla_\nu$, respectively, we obtain the
(non-geodesic) equation of motion of massive test particles as
\begin{eqnarray}  \label{force0}
u^{\nu} \nabla_{\nu} u^{\lambda} =\frac{\mathrm{d}^2x^{\lambda}}{\mathrm{d}%
s^2}+\Gamma_{\mu \nu}^{\lambda }u^{\mu }u^{\nu} = \frac{- h^{\nu \lambda}
\nabla_{\nu} P + h^{\nu \lambda} Q_{\nu}}{\rho + P},  \notag \\
\end{eqnarray}
where
\begin{equation}
h^{\nu \lambda} Q_\nu = \frac{f_T/2}{8 \pi + f_T} h^{\nu \lambda} \nabla_\nu
(\rho - P).
\end{equation}

\subsection{Balance equations in the conformal frame}

In the following for notational simplicity we define first a mix-component
vector field
\begin{eqnarray}
V_{\thickspace \nu}^{\mu} &\equiv& 8\pi \left( T_{\thickspace \nu}^{\mu} -
\frac{\delta_{\thickspace \nu}^{\mu}}{2} T\right) - \frac{\delta_{%
\thickspace \nu}^{\mu}}{2} f -  \notag \\
&&\Bigg[ \left( T_{\thickspace \nu}^{\mu} + \Theta_{\thickspace \nu}^{\mu}
\right) - \frac{\delta_{\thickspace \nu}^{\mu}}{2} \left( T + \Theta \right) %
\Bigg] f_{T}.
\end{eqnarray}
Since $\tilde{\nabla}_\mu \tilde{G}_{\thickspace\nu}^\mu \equiv 0$, then
from Eq.~\eqref{eq:feqc} we at once get the conservation equations in the
conformal frame as
\begin{eqnarray}
\hspace{-0.4cm}&&\tilde{\nabla}_\mu V_{\thickspace\nu}^\mu - 2 V_{\thickspace%
\nu}^\mu \tilde{\nabla}_\mu \ln \left| F \right|  \notag \\
\hspace{-0.4cm} &=& 2 F \tilde{G}_{\thickspace\nu}^\mu \tilde{\nabla}_\mu F
- 2 V_{\thickspace\nu}^\mu \tilde{\nabla}_\mu \ln \left| F \right| = 0
\notag \\
\hspace{-0.4cm} &=& \nabla_\mu V_{\thickspace\nu}^\mu + \left( \tilde{\Gamma}%
_{\lambda\mu}^\mu - \Gamma_{\lambda\mu}^\mu \right) V_{\thickspace%
\nu}^\lambda - \left( \tilde{\Gamma}_{\nu\mu}^\lambda -
\Gamma_{\nu\mu}^\lambda \right) V_{\thickspace\lambda}^\mu  \notag \\
\hspace{-0.4cm} && - 2 V_{\thickspace\nu}^\mu \tilde{\nabla}_\mu \ln \left|
F \right|  \notag \\
\hspace{-0.4cm} &=& \nabla_\mu V_{\thickspace\nu}^\mu + 2 V_{\thickspace%
\nu}^\mu \left( \partial_\mu - \tilde{\nabla}_\mu \right) \ln \left| F
\right| - \frac{V}{2} \partial_\nu \ln \left| F \right|,
\end{eqnarray}
where $V \equiv \delta_{\thickspace\mu}^\nu V_{\thickspace\nu}^\mu = - \left[%
8\pi T-\left(T+\Theta\right)f_T+2f\right] = - F R\left(g,\tilde{\Gamma}%
\right)$ according to Eq.~\eqref{riccsp1}, and we have used Eq.~%
\eqref{eq:Pal-conn}. Taking into account that $F$ is a scalar and $\tilde{%
\nabla}_\mu F = \nabla_\mu F = \partial_\mu F$, then
\begin{equation}
\nabla_\mu V_{\thickspace\nu}^\mu = \frac{V}{2} \partial_\nu \ln \left| F
\right|,
\end{equation}
or equivalently,
\begin{eqnarray}
&&\nabla_\mu \left[8\pi T_{\thickspace\nu}^\mu-\left(T_{\thickspace%
\nu}^\mu+\Theta_{\thickspace\nu}^\mu\right)f_T\right] - \frac{\nabla_\nu}{2} %
\left[F R\left(g,\tilde{\Gamma}\right) - f\right]  \notag \\
&&= - \frac{R\left(g,\tilde{\Gamma}\right) \partial_\nu F}{2},
\end{eqnarray}
where we have used the relation $8\pi T-\left(T+\Theta\right)f_T+f = F
R\left(g,\tilde{\Gamma}\right) - f$ on the left hand side. Hence one can
easily see that the equations above are exactly the same as the energy
balance equations~\eqref{eq:covder-Tmunup1}.

\subsection{Thermodynamic interpretation of $f(R,T)$ gravity theories}

For the sake of completeness we briefly present the thermodynamic
interpretation of $f(R,T)$ gravity theories, as discussed in \cite{Ha14}.
The non-conservation of the matter energy-momentum tensor strongly suggests
that, due to the matter-geometry coupling, particle creation processes may
take place during the cosmological evolution. This phenomenon is also
specific to quantum field theories in curved space-times, as pointed out in
\cite{Parker,Parker1,Parker2}, and it is a consequence of a time varying
gravitational field. Hence, $f(R,T)$ theory, which also involves particle
creation, may lead to the possibility of a semiclassical description of
quantum field theoretical processes in gravitational fields.

\subsubsection{Particle and entropy fluxes, and the creation pressure}

The presence of particles creation implies that the covariant divergence of
the basic equilibrium quantities, including the particle and entropy fluxes,
as well as of the energy-momentum tensor, are now different from zero.
Consequently, all the balance equilibrium equations must be modified to
include particle creation \cite{P-M,Lima,Su}. In the presence of
gravitationally generated matter, the balance equation for the particle flux
$N^{\mu} \equiv nu^{\mu}$, where $n$ is the particle number density, becomes
\begin{equation}
\nabla _{\mu}N^{\mu}=\dot{n}+3Hn=n\Psi,
\end{equation}
where $\Psi $ is the particle production rate, which can be neglected in the case that $\Psi \ll H$. The entropy flux vector is defined to be $S^{\mu} \equiv su^{\mu} = n\sigma u^{\mu}$, where $s$ is the entropy density, and $%
\sigma $ is the entropy per particle. The divergence of the entropy flux
gives
\begin{equation}  \label{62b}
\nabla _{\mu}S^{\mu}=n\dot{\sigma}+n\sigma \Psi\geq 0.
\end{equation}
If we consider a specific $\sigma $ which is a constant, then
\begin{equation}
\nabla _{\mu}S^{\mu}=n\sigma \Psi =s\Psi\geq 0,
\end{equation}
that is, the variation of the entropy is entirely due to (adiabatic gravitational) particle creation processes. Since $s>0$, from the above equation it follows that the particle creation rate must satisfy the condition $\Psi \geq 0$, that is, gravitational fields can generate particles, but the inverse process is prohibited. The
energy-momentum tensor of a fluid in the presence of particle creation must
also be modified to take into account the second law of thermodynamics, so
that \cite{Bar}
\begin{equation}  \label{64}
T^{\mu \nu}=T^{\mu \nu}_\text{eq}+\Delta T^{\mu \nu},
\end{equation}
where $T^{\mu \nu}_\text{eq}$ denotes the equilibrium component~\eqref{eq:perflu-Tmunu}, and $\Delta T^{\mu \nu}$ is the correction due to
particle creation. Due the isotropy and homogeneity of space-time, the extra
contribution to the equilibrium energy-momentum tensor must be represented
by a scalar process. Generally one can write
\begin{equation}
\Delta T_{\; 0}^0=0, \quad \Delta T_{\; i}^j=-P_c\delta_{\; i}^j,
\end{equation}
where $P_c$ is the dynamic \textit{creation pressure} that describes
phenomenologically the thermodynamic effect of particle creation in a
macroscopic system. In a covariant representation we have \cite{Bar}
\begin{equation}
\Delta T^{\mu \nu}=-P_ch^{\mu \nu}=-P_c\left(g^{\mu \nu}+u^{\mu}u^{\nu}\right),
\end{equation}
which immediately gives $u_{\mu}\nabla _{\nu}\Delta T^{\mu \nu}=3HP_c$.
Therefore in the presence of particle creation the total energy balance
equation $u_{\mu}\nabla _{\nu}T^{\mu \nu}=0$, which follows from Eq.~\eqref{64}, immediately gives
\begin{equation}
\dot{\rho}+3H\left(\rho+P+P_c\right)=0.
\end{equation}

The thermodynamic quantities must also satisfy the Gibbs law, which can be
formulated as \cite{Lima}
\begin{equation}
n \mathcal{T} \mathrm{d} \left(\frac{s}{n}\right)=n\mathcal{T}\mathrm{d}\sigma=\mathrm{d}\rho -\frac{\rho+p}{n}\mathrm{d}n,
\end{equation}
where $\mathcal{T}$ is the thermodynamic temperature of the system.

\subsubsection{Thermodynamic quantities in $f(R,T)$ gravity}

After some simple algebraic manipulations the energy balance equation~\eqref{61} can be reformulated as
\begin{equation} \label{76}
\dot{\rho}+3H\left( \rho +P+P_{c}\right) =0,
\end{equation}%
where the creation pressure $P_{c}$ is defined as
\begin{eqnarray}
\hspace{-0.6cm}&&P_{c}=\frac{-f_{T} \left( 1+w \right) \rho}{8\pi +f_{T}}%
\Bigg\{ \ln \left\vert f_{T}\right\vert +\frac{1-w}{2\left( 1+w\right) }-\frac{1}{3H} \times
\notag \\
\hspace{-0.6cm}&&\Bigg[ \nabla _{\mu }\left( u^{\mu }\ln
\left\vert f_{T}\right\vert \right) +\frac{1}{\left( 1+w\right)\rho}\nabla
_{\mu }\left( u^{\mu } \frac{\left( 1-w\right)\rho}{2}\right) \Bigg] %
\Bigg\} , \notag \\
\end{eqnarray}
where we have denoted $w=P/\rho $. Then the generalized balance Eq.~\eqref{61}
can be derived from the divergence of the total energy momentum tensor $%
T^{\mu \nu }$, defined as
\begin{equation}
T^{\mu \nu }=\left( \rho +P+P_{c}\right) u^{\mu }u^{\nu }+\left(
P+P_{c}\right) g^{\mu \nu }.
\end{equation}
On the other hand under the assumption of adiabatic particle production,
with $\dot{\sigma}=0$, the Gibbs law gives
\begin{equation}
\dot{\rho}
=\left(\rho+P\right)\frac{\dot{n}}{n}
=\left(\rho+P\right)\left(\Psi-3H\right),
\end{equation}
which together with the energy balance equation gives immediately the
relation between the particle creation rate and the creation pressure as
\begin{equation}
\Psi=\frac{-3HP_c}{\rho(1+w)}.
\end{equation}
In the framework of the $f(R,T)$ gravity theory we obtain for the particle
creation rate the general expression
\begin{eqnarray}
\hspace{-0.6cm}&&\Psi=\frac{f_{T} }{8\pi +f_{T}}\Bigg\{3H \left[ \ln \left\vert
f_{T}\right\vert + \frac{1-w}{2\left( 1+w\right) } \right] -
\notag\\\hspace{-0.6cm}&& \Bigg[ \nabla _{\mu }\left( u^{\mu }\ln \left\vert
f_{T}\right\vert \right) +\frac{1}{\left( 1+w\right)\rho}\nabla _{\mu
}\left( u^{\mu } \frac{\left( 1-w\right)\rho}{2}\right) \Bigg] \Bigg\} ,
\notag \\
\end{eqnarray}
Hence the condition $\Psi \geq 0$ imposes a strong constraint on the
physical parameters of the theory. In the case of the pressureless dust,
with $P=0$, $w=0$, under the assumption $f_T>0$, we obtain the following
general cosmological constraint that must be satisfied by the function $f_T$
for all times,
\begin{equation}
3H\left(\ln \left|f_T\right|+\frac{1}{2}\right)\geq \nabla
_{\mu}\left(u^{\mu}\ln \left|f_T\right|\right)+\frac{1}{2\rho}\nabla
_{\mu}\left(\rho u^{\mu}\right).
\end{equation}

The divergence of the entropy flux vector can be reformulated in terms of
the creation pressure as
\begin{equation}
\nabla _{\mu}S^{\mu}=\frac{-3 n \sigma H P_c}{\rho (1+w)}.
\end{equation}

Finally, we consider the temperature evolution in a system with particle
creation. In order to fully determine the time behavior of a relativistic
fluid we must add two equations of state for the density and pressure, which
have the general form $\rho =\rho (n, \mathcal{T} )$ and $p=p(n,\mathcal{T})$, respectively. Then
we obtain
\begin{equation}
\dot{\rho}=\left(\frac{\partial \rho }{\partial n} \right)_\mathcal{T}\dot{n}+\left(%
\frac{\partial \rho }{\partial \mathcal{T}} \right)_n\dot{\mathcal{T}}.
\end{equation}
By using the energy and particle balance equations we find
\begin{equation}  \label{78a}
-3H\left(\rho +P+P_c\right)=\left(\frac{\partial \rho }{\partial n}
\right)_\mathcal{T} n\left(\Psi-3H\right)
+\left(\frac{\partial \rho }{\partial \mathcal{T}} \right)_n\dot{\mathcal{T}}.
\end{equation}
With the use of the thermodynamic identity \cite{Bar}
\begin{equation}
\mathcal{T}\left(\frac{\partial P}{\partial \mathcal{T}}\right)_n=\rho+P-n\left(\frac{\partial
\rho}{\partial n}\right)_\mathcal{T},
\end{equation}
Eq.~\eqref{78a} yields the temperature evolution of a relativistic fluid in
the presence of matter creation as
\begin{equation}
\frac{\dot{\mathcal{T}}}{\mathcal{T}}=\left(\frac{\partial P}{\partial \rho}\right)_n\frac{\dot{n}}{n}.
\end{equation}
In the particular case $\left(\partial P/\partial \rho\right)_n=w=\mathrm{
constant}$, we obtain for the temperature-particle number dependence the
simple expression $\mathcal{T} \sim n^w$.

\subsubsection{The case $w=-1$}

Based on the homogeneous and isotropic Friedmann-Robertson-Walker metric, and on the energy conservation equation $ \dot{\rho} + 3H(1+w)\rho = 0$, in \cite{r3s}, general cosmological thermodynamic properties with an \textit{arbitrary, varying equation-of-state parameter $w(a)$}, where $a$ is the scale factor, were discussed. The $w=-1$-crossing problem of $w$ was explicitly pointed out, and the  behaviors of the quantities ($\rho(a)$, $\mu(a)$, $\mathcal{T}(a)$, etc.) at/near $w=-1$ were discussed. As a result of this study it was concluded that \textit{all cosmological quantities must be, and indeed they are regular and well-defined for all values of $w(a)$} \cite{r3s}. In the present thermodynamical approach we have assumed that matter is created in an ordinary form, and therefore all our previous results are valid for $w\geq 0$. However, the thermodynamic approach and interpretation of the $f(R,T)$ gravity can be extended to the case $w<0$. In the following we will consider this problem, and we show that our results are valid, in the sense of regularity and well-definiteness even in the case of $w=-1$. In particular, we concentrate on the temperature evolution equation,
\begin{equation} \label{tempevol}
\frac{\dot{\mathcal{T}}}{\mathcal{T}}
=\left(\frac{\partial P}{\partial \rho}\right)_n\frac{\dot{n}}{n}.
\end{equation}
which still holds even if $w= P/\rho = -1$. The demonstration is as follows. First we consider the perfect-fluid energy-momentum balance equation
\begin{eqnarray} \label{61}
&&\dot{\rho}+3(\rho + P) H = \frac{-f_T}{8\pi +f_T} \bigg[(\rho + P) u^\mu \nabla_\mu \ln |f_T|
\notag\\&& + u^\mu \nabla_\mu (\frac{\rho - P}{2}) \bigg], \quad f = f (R(g,\tilde{\Gamma}), T).
\end{eqnarray}
When $w=-1$, Eq.~\eqref{61} becomes
\begin{equation}
  \dot{\rho} = \frac{-f_T}{8\pi +f_T} u^\mu \nabla_\mu \rho
  \equiv -3H P_c,
\end{equation}
where $P_c$ is the matter creation pressure. Under the assumption of adiabatic particle production,
with $\dot{\sigma}=0$, the Gibbs law gives
\begin{equation}
\dot{\rho} = (\rho+P)\frac{\dot{n}}{n} = 0,
\end{equation}
where $\sigma$ is the entropy per particle. That is, from the above two equations, we obtain
\begin{equation}
  \dot{\rho} = P_c = 0.
\end{equation}
Since $\rho = \rho \left(n, \mathcal{T}\right)$, we have
\begin{equation}
\dot{\rho}=\left(\frac{\partial \rho }{\partial n}\right)_\mathcal{T} \dot{n} +\left(\frac{\partial \rho}{\partial \mathcal{T}}\right)_n\dot{\mathcal{T}}
= 0.
\end{equation}
Combining the above equation and the thermodynamic identity \cite{Bar},
\begin{equation}
\mathcal{T}\left(\frac{\partial P}{\partial \mathcal{T}}\right)_n
= \rho+P-n\left(\frac{\partial\rho}{\partial n}\right)_\mathcal{T}
= -n\left(\frac{\partial\rho}{\partial n}\right)_\mathcal{T},
\end{equation}
it immediately follows that Eq.~\eqref{tempevol} still holds even for negative values of $w=-1$. If $w=-1=\text{constant}$, we can obtain from Eq.~\eqref{tempevol} that $n\mathcal{T}$ is a constant, or $\mathcal{T}\sim 1/n$. This relation shows that for very low density "dark energy" particles their thermodynamic temperature is extremely high, while high particle number (density) systems have a very low temperature. In the limit $n\rightarrow \infty$, the temperature of the "dark energy" made system tends to zero.

\section{Cosmology of Palatini $f(R,T)$ gravity}\label{sect3}

In the present Section we investigate the cosmological implications of the
Palatini formulation of $f(R,T)$ gravity. We assume that the Universe is
flat, homogeneous and isotropic, with the metric given in comoving
coordinates by the Friedmann-Robertson-Walker metric,
\begin{equation}
d s^2 = g_{\mu\nu} \mathrm{d} x^\mu \mathrm{d} x^\nu = - dt^2 + a^2 (t)\left(dx^2+dy^2+dz^2\right),  \label{eq:FRW-metric-flat}
\end{equation}
where $a(t)$ is the scale factor. We also introduce the Hubble function,
defined as $H=\dot{a}/a$. We assume that the matter content of the Universe
consists of a perfect fluid, that can be characterized by two thermodynamic
parameters, the energy density $\rho $, and the pressure $P$, respectively.
As for the relations of the geometric quantities in the $g$ and $\tilde{g}$
frames, their detailed computation is presented in Appendix~\ref{app3}.

\subsection{Generalized Friedmann equations in Palatini $f(R,T)$ gravity}

Substituting the expression of the perfect-fluid energy-momentum tensor as
well as Eqs.~\eqref{eq:Thetamunu2}, \eqref{eq:perflu-T&theta} and \eqref{G00}
into the Palatini field equations~\eqref{eq:feqp1}, from the 00 component we
obtain the first modified Friedmann equation as
\begin{eqnarray}  \label{fRT-frideq1}
&&3 H^2 = \frac{1}{2 F} \bigg[ 8 \pi \left( \rho + 3 P \right) + (\rho + P)
f_T  \notag \\
&& + f - \frac{3{\dot{F}}^2}{2 F} - 6 H \dot{F} \bigg].
\end{eqnarray}
Similarly, with the help of Eqs.~\eqref{eq:Thetamunu2}, %
\eqref{eq:perflu-T&theta} and \eqref{Gii}, we can obtain the second modified
Friedmann equation from the `$ii$' components of Eq.~\eqref{eq:feqp1},
\begin{eqnarray}  \label{fRT-frideq2}
&& 2 \dot{H} + 3 H^2 = \frac{-1}{2 F} \bigg[ 8 \pi \left(
\rho - P \right) + (\rho + P) f_T  \notag \\
&& - f - \frac{3{\dot{F}}^2}{2 F} + 2 \ddot{F} + 4 H \dot{F} \bigg].
\end{eqnarray}
Note that the first modified Friedmann equation~\eqref{fRT-frideq1} can be
written more compactly as
\begin{equation}  \label{fRT-frideq11}
\left(H + \frac{\dot{F}}{2 F} \right)^2 = \frac{8 \pi \left( \rho + 3 P
\right) + (\rho + P) f_T + f}{6 F}.
\end{equation}
Finally, we substitute Eqs.~\eqref{eq:Thetamunu2} and \eqref{eq:Ricciscalar3}
into Eq.~\eqref{feqp-tr} and obtain the trace equation
\begin{eqnarray}  \label{fRT-frideq-tr}
\hspace{-0.6cm}&& F R\left(g,\tilde{\Gamma}\right) - 2f = 8 \pi T - \left( T
+ \Theta \right) f_{T}=  \notag \\
\hspace{-0.4cm}&& 6 F \left( \dot H + 2 H^2 \right) - 3 \left( \frac{{\dot{F}%
}^2}{2 F} - \ddot F - 3 H \dot F \right) - 2f.
\end{eqnarray}

By eliminating the term $3H^{2}$ between the two generalized Friedmann
equations we obtain the evolution equation for the Hubble function as given
by
\begin{equation}
\dot{H}=\frac{-1}{2F}\left[ \left( 8\pi +f_{T} \right) \left( \rho +P\right) - \frac{3\dot{F}^{2}}{2F} + \ddot{F} - H\dot{F} \right].  \label{Hdot}
\end{equation}

The $f\left(R\left(g,\tilde{\Gamma}\right), T\right) \rightarrow
f\left(R\left(g,\tilde{\Gamma}\right)\right)$ limits of the two modified
Friedmann equations can be given by
\begin{equation}
6 H^2 F - f = 8 \pi \left( \rho + 3 P \right) - \frac{3{\dot{F}}^2}{2 F} - 6
H \dot{F}
\end{equation}
and
\begin{equation}
- 2 F \left( 2 \dot{H} + 3 H^2 \right) + f = 8 \pi \left( \rho - P \right) -
\frac{3{\dot{F}}^2}{2 F} + 2 \ddot{F} + 4 H \dot{F}.
\end{equation}
If we go one step further and take $f\left(R\left(g,\tilde{\Gamma}%
\right)\right) \rightarrow R\left(g,\tilde{\Gamma}\right)$, then $f = R (g)
= 6 \left( \dot H + 2 H^2 \right)$ and the above two equations reduce to
\begin{equation}  \label{fR-frideq1}
6 H^2 - 6 \left( \dot H + 2 H^2 \right) = - 6 \left( \dot H + H^2 \right) =
8 \pi \left( \rho + 3 P \right)
\end{equation}
and
\begin{equation}  \label{fR-frideq2}
- 2 \left( 2 \dot{H} + 3 H^2 \right) + 6 \left( \dot H + 2 H^2 \right) = 2
\dot H + 6 H^2 = 8 \pi \left( \rho - P \right).
\end{equation}
That is,
\begin{equation}  \label{R-frideq2}
\dot H + H^2 = - \frac{4 \pi}{3} \left( \rho + 3 P \right),
\end{equation}
and
\begin{equation}  \label{R-frideq1}
H^2 = \frac{8 \pi}{3} \rho.
\end{equation}
Hence the first modified Palatini $f(R,T)$ Friedmann equation~%
\eqref{fRT-frideq1} reduces to the ordinary second Friedmann equation~%
\eqref{R-frideq2} when $f(R, T) \rightarrow R$.

\subsection{The energy balance equation}

With the help of Eq.~\eqref{eq:non0-cons}, we can directly work out the
covariant divergence of the energy-momentum tensor~\eqref{eq:perflu-Tmunu}
as
\begin{eqnarray}
\nabla_\mu T_{\thickspace i}^\mu &=& \partial_\mu T_{\thickspace i}^\mu +
\Gamma_{\thickspace \mu \nu}^\mu T_{\thickspace i}^\nu - \Gamma_{\thickspace %
\mu i}^\nu T_{\thickspace \nu}^\mu  \notag \\
&=& \partial_i T_{\thickspace i}^i + \Gamma_{\thickspace \mu i}^\mu T_{%
\thickspace i}^i - \Gamma_{\thickspace 0 i}^0 T_{\thickspace 0}^0 - 3
\Gamma_{\thickspace 1 1}^1 T_{\thickspace 1}^1  \notag \\
&=& 0, \quad i = 1, 2, 3,  \label{eq:covderiv-Tmu1}
\end{eqnarray}
and
\begin{eqnarray}
\nabla_\mu T_{\thickspace 0}^\mu &=& \partial_\mu T_{\thickspace 0}^\mu +
\Gamma_{\thickspace \mu \nu}^\mu T_{\thickspace 0}^\nu - \Gamma_{\thickspace %
\mu 0}^\nu T_{\thickspace \nu}^\mu  \notag \\
&=& \partial_0 T_{\thickspace 0}^0 + \Gamma_{\thickspace \mu 0}^\mu T_{%
\thickspace 0}^0 - \Gamma_{\thickspace 0 0}^0 T_{\thickspace 0}^0 - 3
\Gamma_{\thickspace 1 0}^1 T_{\thickspace 1}^1  \notag \\
&=& - \dot \rho - 3 H (\rho + P).  \label{eq:covderiv-Tmu0}
\end{eqnarray}
Substituting the above two equations, as well as Eqs.~\eqref{eq:Thetamunu2}
and \eqref{eq:perflu-T&theta}, into the already known Palatini $f(R,T)$
energy balance equation~\eqref{61},
\begin{equation}
\nabla_{\mu} T_{\thickspace \nu}^\mu = \frac{f_{T}}{8\pi - f_{T}} \bigg[ %
\left( T_{\thickspace \nu}^\mu + \Theta_{\thickspace \nu}^\mu \right)
\nabla_{\mu} \ln \left| f_{T} \right| + \nabla_{\mu} \Theta_{\thickspace %
\nu}^\mu - \frac{1}{2} \nabla_{\nu}T \bigg],
\end{equation}
we obtain
\begin{equation}
\dot{\rho} + 3 H (\rho + P) = \frac{f_T}{8 \pi} \left[ \frac{\dot{P} - 3
\dot{\rho}}{2} - \left( 3 H + \frac{\dot{f_T}}{f_T} \right) (\rho + P) %
\right].  \label{eq:enerconlike}
\end{equation}
When $f(R,T) \rightarrow f(R)$, the energy balance equation reduces to the
ordinary conservation equation
\begin{equation}
\dot{\rho} + 3 H (\rho + P) = 0.
\end{equation}

\subsection{Deceleration parameter and equation of state of the Universe}

An important cosmological parameter, indicating the
accelerating/decelerating nature of the cosmological dynamics is the
deceleration parameter $q$ defined to be
\begin{equation}
q=\frac{\mathrm{d}}{\mathrm{d} t} \left( \frac{1}{H} \right) - 1
=-\frac{\dot{H}}{H^{2}}-1.
\end{equation}
Using Eqs.~\eqref{fRT-frideq1} and \eqref{Hdot}, we immediately obtain
\begin{equation}
q=3\frac{\left( 8\pi +f_{T}\right) \left( \rho + P\right) - \frac{3\dot{F}^{2}}{2F} + \ddot{F}-H\dot{F}}{8\pi \left( \rho
+3P\right) +(\rho +P)f_{T}+f-\frac{3{\dot{F}}^{2}}{2F}-6H\dot{F}}-1.
\end{equation}

For a vacuum Universe with $\rho =P=0$, the condition for an accelerated
expansion $q<0$ reduces to
\begin{equation}\label{qP}
\frac{\ddot{F}-3\dot{F}^{2}/2F-H\dot{F}}{f-3{\dot{F}%
}^{2}/2F-6H\dot{F}}< \frac{1}{3}.
\end{equation}

The deceleration parameter can also be defined, by analogy with the standard
general relativistic cosmology, in terms of the effective parameter $w_\text{eff}$ of the equation of state of the Universe as
\begin{equation}
q=\frac{1+3w_\text{eff}}{2},
\end{equation}%
giving
\begin{equation}
w_\text{eff}=\frac{2q-1}{3}.
\end{equation}
By using the above definition we obtain for the effective parameter of the
equation of state of the Universe the expression
\begin{equation}
w_\text{eff}=2\frac{ \left( 8\pi +f_{T}\right) \left( \rho +P\right)-\frac{3\dot{F}^{2}}{2F}+\ddot{F}-H\dot{F} }{8\pi \left( \rho +3P\right) +(\rho
+P)f_{T}+f-\frac{3{\dot{F}}^{2}}{2F}-6H\dot{F}}-1.
\end{equation}

\subsection{The de Sitter solution}

Next, we investigate the possibility of the existence of a de Sitter type
solution in the framework of the Palatini formulation of $f(R,T)$ gravity.
The de Sitter solution corresponds to $H=H_{0}=\mathrm{constant}$, and $\dot{%
H}=0$, respectively. Assuming that the Universe is filled with a
pressureless dust, we have $P=0$, and $T=-\rho$. Moreover, we adopt for the
function $f$ the functional form $f(R,T)=k(R)+g(T)$, where for simplicity we
take $g(T)=8 \pi \beta T$, with $\beta$ a constant. Then the energy
balance Eq.~\eqref{eq:enerconlike} takes the form
\begin{equation}
\left(1+ \frac{3\beta}{2} \right)\dot{\rho}=-3H_0\left(1+ \beta \right)\rho,
\end{equation}
with the general solution given by
\begin{equation}
\rho (t)=\rho _0e^{-\tilde{\alpha} t},
\end{equation}
where $\rho _0$ is an integration constant, and we have denoted
\begin{equation}
\tilde{\alpha }=\frac{3H_0\left(1+ \beta \right)}{1+3\beta/2}.
\end{equation}
Eq.~\eqref{Hdot} becomes
\begin{equation}  \label{114}
\ddot{F}-\frac{3\dot{F}^2}{2F}-H_0\dot{F}+8\pi\left(1+ \beta \right)\rho_0e^{-\tilde{\alpha }t}=0.
\end{equation}

In the limit of large times the last, exponential term in the above equation
becomes negligibly small, and hence we can approximate the solution of Eq.~\eqref{114} as
\begin{equation}
F(t)\approx \frac{F_0}{\left(e^{H_0 t}-1\right)^2},
\end{equation}
where $F_0$ is an arbitrary constant of integration, and, without any loss
of generality we have taken the second integration constant as zero. Then
the first generalized Friedmann equation \eqref{fRT-frideq1} gives the
Lagrangian function of the model as
\begin{equation}  \label{116}
f(t)=6H_0^2F+\frac{3\dot{F}^2}{2F}+6H_0\dot{F}+8\pi\left(1+ \beta \right)T.
\end{equation}
On the other hand the trace equation \eqref{fRT-frideq-tr} gives (note that in the following $R=R(g)$)
\begin{equation}  \label{117}
FR+3\left(\frac{\dot{F}^2}{F}+4H_0\dot{F}\right)=8\pi\left(1+ \beta \right)T+2f,
\end{equation}
or
\begin{equation}  \label{118}
R(t)=12H_0^2+24\pi\left(1+ \beta \right)\frac{T}{F}.
\end{equation}

Eqs.~\eqref{116} and \eqref{118} give a parametric representation of $f$ as
a function of $R$, with $t$ taken as parameter. Once the function $t=t(R)$
is obtained from Eq.~\eqref{118}, by substituting it in Eq.~\eqref{116} we can find the explicit dependence of $f$ on $R$.

\subsection{Comparison with the metric $f(R,T)$ cosmology}

Using the Friedmann - Robertson - Walker metric, and the field equations (\ref{eq:feqm1}),
 we can similarly derive the two Friedmann equations in the metric formulation.  The cosmological equations in the metric $f(R,T)$ gravity are different from their counterparts in the Palatini formalism, due to the presence of some dynamical terms related to $f_R$, and they are given by
\begin{equation}
3 H^2 = \frac{1}{2 F} \bigg[ 8 \pi \left( \rho + 3 P \right) + (\rho + P)f_T + f + 3 \ddot{f_R}  + 3 H \dot{f_R} \bigg].
\end{equation}
\bea
2 \dot{H} + 3 H^2 &=& \frac{-1}{2 F} \bigg[ 8 \pi \left(
\rho - P \right) + (\rho + P) f_T- \nonumber\\
&&f - \ddot{f_R} - 5 H \dot{f_R} \bigg].
\eea
Besides, the deceleration parameter $q = -\dot{H}/H^2 -1$ can be obtained as
\begin{equation}
q=3 \frac{\left( 8\pi +f_T \right) \left( \rho + P\right) + \ddot{f_R} - H\dot{f_R}}{8\pi \left( \rho
+3P\right) +(\rho +P)f_T + f + 3 \ddot{f_R}+3H\dot{f_R}}-1.
\end{equation}

For $\rho=P=0$, an accelerated Universe with $q<0$ requires that
\begin{equation}
\frac{\ddot{f_R}-H\dot{f_R}}{f+ 3 \ddot{f_R}+3 H\dot{f_R}}< \frac{1}{3}.
\end{equation}

The condition for accelerated expansion in the metric $f(R,T)$ gravity is very different from the similar condition, given by Eq.~(\ref{qP}), in the Palatini formulation of the theory. The presence of the extra term $3\dot{F}^2/2F$ in Eq.~(\ref{qP}) may have a significant effect on the transition from the decelerating to the accelerating phase. In the Palatini formulation the moment of the transition to the accelerated expansion with $q\leq 0$ is determined by the equation
\be
\ddot{f_R}-\dot{f}_R^2/f_R+H\dot{f}_R-f/3=0,
\ee
with $f_R=f_R\left(R\left(g,\tilde{\Gamma}\right),T\right)$, while in the metric $f(R,T)$ gravity the same condition is given by the much simpler expression
\be
6H\dot{f}_R+f=0,
\ee
with $f=f\left(R(g),T\right)$. It is interesting to note that the conditions for the transition to an accelerated expansion in the vacuum case are independent in both approaches on $f_T$. However, the Palatini formulation of $f(R,T)$ gravity allows a much richer cosmological dynamics, as compared to the metric formulation.

As for the energy-momentum balance equation and the non-geodesic equation of motion of massive test particles, since they are all derived from the divergence of the energy-momentum tensor, and since $\nabla_\mu T_{\;\nu}^\mu$ is the same in the two formalisms, independently of the functional form of $f$, the energy balance equations and the equations of motion have the same functional form in both approaches.

\subsection{Specific cosmological models in the Palatini $f(R,T)$ gravity}

In the present Section we will investigate two specific cosmological models
in the framework of the Palatini formulation of $f(R,T)$ gravity. We will
assume that the action of the gravitational field has the general form
\begin{equation}
f(R,T)=R\left(g,\tilde{\Gamma}\right)+k\left(R\left(g,\tilde{\Gamma}\right) \right)+8\pi
\beta g(T),
\end{equation}
where $\beta $ is a constant, and $k \left(R\left(g,\tilde{\Gamma}\right) \right)$ and $g(T)$ are arbitrary functions of the Ricci scalar and the trace of the
matter energy-momentum tensor. For the function $g(T)$ for simplicity we
will assume a simple linear dependence on $T$, so that $g(T)=T$. As for the
function $k\left(R\left(g,\tilde{\Gamma}\right) \right)$, we will consider two cases,
corresponding to the Starobinsky model $k(R)\sim R^2\left(g,\tilde{\Gamma}%
\right)$ \cite{Star}, and the $1/R\left(g, \tilde{\Gamma}\right)$ case,
respectively.

\subsubsection{$f(R,T) = R\left(g,\tilde{\Gamma}\right) + \frac{%
\alpha}{16 \pi} R^2\left(g,\tilde{\Gamma}\right) + 8\pi
\protect\beta g(T)$}

We consider a Palatini $f(R,T)$ model, specified by a functional form of $f(R,T)$ given as
\begin{equation}
f(R,T) = R\left(g,\tilde{\Gamma}\right) + \frac{\alpha}{16 \pi} R^2\left(g,%
\tilde{\Gamma}\right) + 8 \pi \beta g(T),
\end{equation}
where $\alpha$, $\beta$ are constants, $g(T) $ is a function that depends on
$T$ solely, and for simplicity we set $g(T) = T$. For this Lagrangian we
immediately find $F = f_R = 1 + \left(\alpha/8 \pi\right) R\left(g,\tilde{%
\Gamma}\right)$, $f_T = 8 \pi \beta g_T$, $g_T \equiv \partial g(T)/\partial
T=8\pi \beta$. Moreover, we assume that the cosmological fluid satisfies a
linear barotropic equation of state of the form $P=w\rho$, $w=\mathrm{%
constant}$.

Consequently, from the trace~\eqref{riccsp1} of the Palatini field equations
we first obtain
\begin{equation}
\left(1+\frac{\alpha R}{8\pi}\right)R =8\pi T-8\pi \beta
\left(T+\Theta\right)+2f,
\end{equation}
thus obtaining
\begin{equation}  \label{riccism1}
R= 8 \pi\left[ (1 - 3w) + \beta (3 - 5w) \right] \rho \equiv 8 \pi \beta_{0}
\rho,
\end{equation}
When $\beta \rightarrow 0$, $\beta_0 \rightarrow 1 - 3w$, and $R\left(g,%
\tilde{\Gamma}\right) \rightarrow - 8 \pi T$, respectively. Substituting Eq.~%
\eqref{riccism1} back into the expressions of $f(R,T)$ and $F$, we obtain
\begin{equation}  \label{fm1}
f = 8 \pi \left[ \beta_{0} + \frac{{\beta_{0}}^2}{2}\alpha\rho + \beta (3 w
- 1) \right] \rho
\end{equation}
and
\begin{equation}  \label{Fm1}
F = 1 + \beta_{0} \alpha \rho.
\end{equation}
Substituting Eqs.~\eqref{riccism1}--\eqref{Fm1} into the Friedmann equation~\eqref{fRT-frideq11}, we find
\begin{equation}  \label{feq1m1}
\left[ H + \frac{\beta_0 \alpha \dot{\rho}}{2 \left(1 + \beta_0 \alpha \rho
\right)} \right]^2 = \frac{1 + \frac{\beta}{2} (3-w) + \frac{{\beta_{0}}^2}{4%
}\alpha\rho}{\left(1 + \beta_0 \alpha \rho \right)} \frac{8 \pi \rho}{3}.
\end{equation}
Substitution of the expression of $f_{T}$ into the balance equation~%
\eqref{eq:enerconlike} gives
\begin{equation}
\dot{\rho}=-\beta _{1}H\rho ,  \label{beqm1}
\end{equation}%
where we have denoted
\begin{equation}
\beta _{1}\equiv \frac{3(1+\beta )(1+w)}{1+\frac{\beta }{2}(3-w)}.
\end{equation}%
When $\beta \rightarrow 0$, then $\beta_{1}=3(1+w)$. Substituting
Eq.~\eqref{beqm1} into Eq.~\eqref{feq1m1}, we find
\begin{eqnarray} \label{feq1m12}
&& \frac{\left[ 1+\beta_{0}\left( 1-\frac{\beta_{1}}{2}\right)
\alpha \rho \right]^{2}}{1+\beta _{0}\alpha \rho }H^{2} =
\notag\\&&
\frac{8\pi \rho }{3}\left[ 1+\frac{\beta }{2}(3-w)
+\frac{{\beta_{0}}^{2}}{4}\alpha \rho \right].
\end{eqnarray}

Taking into account the limit $\alpha\rho \rightarrow 0$, we have the series expansion
\begin{equation}
\frac{ \left(1+\beta_{0}\alpha \rho\right) \left[ 1+\frac{\beta}{2} (3-w) + \frac{{\beta_{0}^{2}}}{4}\alpha\rho \right]}{\left[ 1 + \beta_{0} \left( 1-\frac{\beta_{1}}{2} \right) \alpha \rho \right]^{2}}
\approx \beta_{2} + \beta_{3}\alpha\rho,
\end{equation}
where we have denoted $\beta_{2}=1+\beta(3-w)/2$ and $\beta_{3} = \beta_{0} \left\{ \beta_{0}/4 - \left[1 + \beta (3-w)/2 \right]\left(1-\beta _{1}\right) \right\}$. Then Eq.~\eqref{feq1m12} takes the form
\begin{equation}
H^{2}=\frac{8\pi \rho }{3}\left( \beta_{2} + \beta_{3} \alpha \rho \right) .
\label{Heq1}
\end{equation}
Eq.~\eqref{beqm1} can be immediately integrated to give
\begin{equation}
\rho = \rho_{0} a^{- \beta_{1}}.
\end{equation}
Hence Eq.~\eqref{Heq1} becomes a first order differential equation,
\begin{equation}
\dot{a}=\sqrt{\frac{8\pi }{3}\rho _{0}}a^{1-\beta _{1}}\sqrt{\beta_{2} a^{\beta_{1}}+\beta_{3} \alpha \rho_{0}},
\end{equation}
with the general solution given by
\begin{equation}
a(t)=\left( \frac{2\pi}{3} {\beta_{1}}^{2} \beta_{2} \rho_{0} t^{2}-\frac{\beta_{3}}{\beta_{2}} \alpha \rho_0 \right)^{\frac{1}{\text{$\beta $}_{1}}}.
\end{equation}

In the limit of large times we obtain $a(t) \sim t^{2/\beta _{1}}$,
and $H(t)=\left( 2/\beta _{1}\right) \left( 1/t\right) $, respectively. The
deceleration parameter in this model is given by $q=\beta _{1}/2-1$, and
once the coefficient $\beta _{1}$ satisfies the condition $\beta _{1}/2<1$,
the Universe will experience an accelerated evolution. For an arbitrary $w$,
the condition for a power law type accelerated expansion is $\beta <-\left(
1+3w\right) /4w$, a condition which shows that for $w>0$, $\beta $ must take
negative values.

\subsubsection{$f(R,T)=R\left(g,\tilde{\Gamma}\right)-\frac{\alpha^2%
}{3 R\left(g,\tilde{\Gamma}\right)}+8\pi \beta g(T)$}

Now let us consider the following $f(R,T)$ gravity Palatini type model
\begin{equation}  \label{eq:fofinR}
f\left(R\left(g,\tilde{\Gamma}\right),T\right)=R\left(g,\tilde{\Gamma}%
\right)-\frac{\alpha^2}{3 R\left(g,\tilde{\Gamma}\right)}+8\pi \beta g(T),
\end{equation}
which immediately gives $F=1+\alpha^2 R^{-2}\left(g,\tilde{\Gamma}\right) /3$%
.

From the trace~\eqref{riccsp1} of the Palatini field equations we obtain
\begin{eqnarray}  \label{eq:feqtraceinR}
R^2\left(g,\tilde{\Gamma}\right) + \Phi R \left( g,\tilde{\Gamma} \right) -
\alpha^2 = 0,
\end{eqnarray}
where we have denoted
\begin{equation}
\Phi= 8 \pi \left[ T - (T + \Theta) \beta g_T + 2 \beta g (T) \right].
\end{equation}
The algebraic equation Eq.~\eqref{eq:feqtraceinR} has two distinct
solutions. However, only one of them can be adopted as the physical
solution, more exactly the one which under the limit $f \rightarrow R $,
would give $R=-8\pi T$, which is the trace of Einstein field equation. Hence
when $\Phi \leq 0$, the physical solution of Eq.~\eqref{eq:feqtraceinR} is
\cite{1/R}
\begin{eqnarray}  \label{eq:approxR}
R\left(g,\tilde{\Gamma}\right) = -\frac{1}{2}\left( \Phi-\sqrt{\Phi^2 +4
\alpha^2} \right).
\end{eqnarray}

In the following we will study this cosmological model under the
approximation $\alpha \gg \left| \Phi \right|$ \cite{1/R}, which allows us
to find the evolution of the Universe at later times when $R \left(g,\tilde{%
\Gamma}\right)$ is relatively small. Under the adopted approximation we can
expand $R\left(g,\tilde{\Gamma}\right)$ to the order of $O\left( \lvert \Phi
\rvert \right)$, and obtain
\begin{eqnarray}
R\left(g,\tilde{\Gamma}\right) &\approx & -\frac{1}{2}\Phi +\alpha,
\label{eq:approx1} \\
f\left(R\left(g,\tilde{\Gamma}\right), T \right) & \approx & -\frac{2}{3}%
\Phi + \frac{2}{3} \alpha + 8 \pi \beta g(T),  \label{eq:approx2} \\
F &\approx & \frac{1}{3 \alpha} \Phi + \frac{4}{3}.  \label{eq:approx3}
\end{eqnarray}

For simplicity, we set again $g(T)=T$. Using the approximations~%
\eqref{eq:approx1}, \eqref{eq:approx2} and \eqref{eq:approx3}, respectively,
we can easily obtain the Palatini $f(R,T)$ field equations for this model as
\begin{eqnarray}
&&\tilde{R}_{\mu\nu}
= \frac{g_{\mu \nu}}{4}\alpha + 8 \pi \left( \frac{3}{4} T_{\mu \nu} -\frac{5g_{\mu \nu}}{16}T \right) - 8 \pi \beta \times
\notag\\&& \left[ \left( \frac{3}{4} \Theta_{\mu \nu} -\frac{5g_{\mu
\nu}}{16} \Theta \right) + \left( \frac{3}{4} T_{\mu \nu} - \frac{g_{\mu \nu}
}{16}T\right) \right].
\end{eqnarray}
When $f(R,T) \rightarrow f(R)$, $\beta \rightarrow 0$, and the above field
equations reduce to the field equations considered within the Palatini
formulation of the $f(R)$ gravity, considered in \cite{1/R} \footnote{%
Note that in our result there is a minus sign before $\alpha$, when compared
with Eq. (24) in \cite{1/R}. The reason is that a different solution has
been chosen in Eq. (17) of \cite{1/R}, while we have adopted the solution
given by our Eq.~\eqref{eq:approxR}.},
\begin{equation}
\tilde{R}_{\mu\nu}
= \frac{g_{\mu \nu}}{4} \alpha + 8 \pi \left( \frac{3}{4} T_{\mu\nu} - \frac{5g_{\mu \nu}}{16}T \right).
\end{equation}

With the equation of state $w = P/\rho$, and since $T=-\rho + 3 P$ for a
perfect fluid, we have $\Phi = -8 \pi \beta_0 \rho$; besides, we already
know from Eq.~\eqref{beqm1} that $\dot{\rho}=-\beta_1 H \rho $. Thus,
similarly to Eq.~\eqref{feq1m12}, we have
\begin{eqnarray}
&&\frac{\left[1 - \pi \beta_0 \left( 2 - \beta_1 \right) \frac{\rho}{\alpha} %
\right]^2}{1-2\pi \beta_0 \frac{\rho}{\alpha}} H^2 =  \notag \\
&& \left[1+\left(3 +4\beta\right) w +\frac{2}{3}\beta_0 \right]\pi \rho +
\frac{\alpha}{12}.
\end{eqnarray}

In the first order approximation in $\rho/\alpha$  we obtain
\begin{equation}
H^2=\frac{\alpha}{12} + \left[ \frac{1}{6} \beta_0 (5-\beta_1) + (3+4 \beta) w+1\right] \pi \rho .
\end{equation}

Hence in the present model a cosmological constant $\alpha /12$ is
automatically generated, due to the $1/R$ modification of the gravitational
Lagrangian. Hence for $\rho \rightarrow 0$, the Universe will end in a de
Sitter phase, with $H=H_{0}=\sqrt{\alpha /12}=\mathrm{constant}$. For $\rho
\neq 0$, we have $\rho =\rho_{0} a^{-\beta_{1}}$, and the evolution of the
scale factor is determined by the equation
\begin{equation}
\dot{a}=a\sqrt{\frac{\alpha }{12}+\frac{\beta_{4}\rho _{0}}{a^{\beta _{1}}}%
},
\end{equation}%
where we have denoted $\beta_{4}= \left[ \beta _{0}(5-\beta_{1})/6+(3+4\beta)w+1\right]\pi$, with the general solution given by
\begin{eqnarray}
a(t)&=& 12^{\frac{1}{\beta _1}} \Bigg\{\sqrt{\frac{\beta _4 \rho _0}{\alpha }} \sinh \Bigg[\sinh
   ^{-1}\left(\frac{a_0^{\frac{\beta _1}{2}}}{2 \sqrt{3} \sqrt{\frac{\beta _4 \rho
   _0}{\alpha }}}\right)+\nonumber\\
&&   \frac{\sqrt{\alpha } \beta _1 \left(t-t_0\right)}{4
   \sqrt{3}}\Bigg]\Bigg\}^{\frac{2}{\beta _1}},
\end{eqnarray}
where we have used the initial condition $a\left( t_{0}\right) =a_{0}$.

For the Hubble function we obtain
\begin{eqnarray}
H(t) &=&\frac{\sqrt{\alpha }}{2 \sqrt{3}} \coth \Bigg[\frac{\sqrt{\alpha } \beta _1
   \left(t-t_0\right)}{4 \sqrt{3}}+\nonumber\\
&&   \sinh ^{-1}\left(\frac{a_0^{\beta _1/2}}{
   \sqrt{12\beta _4 \rho _0/\alpha }}\right)\Bigg],
\end{eqnarray}%
while the deceleration parameter of this model is given by
\bea
q(t)&=&\frac{1}{2} \beta _1 \text{sech}^2\Bigg[\frac{\sqrt{\alpha } \beta _1
   \left(t-t_0\right)}{4 \sqrt{3}}+\nonumber\\
&&   \sinh ^{-1}\Bigg(\frac{a_0^{\beta _1/2}}{
    \sqrt{12\beta _4 \rho _0/\alpha }}\Bigg)\Bigg]-1,
\eea
where $\mathrm{sech} \; t = 1/\mathrm{cosh} \; t$. In the limit  $t\rightarrow \infty$, $q\rightarrow-1$, and hence the Universe ends in a de Sitter type accelerating phase, independently of the matter equation of state.

\paragraph{The matter dominated phase}

The matter dominated phase corresponds to the choice $w=0$ in the matter equation of state, that is, to a Universe filled with pressureless baryonic matter. In order to investigate the behavior of the cosmological model during matter domination, we consider the series expansion of the cosmological parameters. Thus we obtain
\begin{eqnarray}
a(t)&\approx& a_{0}+\frac{a_{0}^{1-\frac{\beta _{1}}{2}}\beta _{4}\rho _{0}%
\sqrt{\frac{\alpha a_{0}^{\beta _{1}}}{\beta _{4}\rho _{0}}+12}}{2\sqrt{3}}%
\left( t-t_{0}\right) +\nonumber\\
&&\frac{1}{24}a_{0}^{1-\beta _{1}}\left[ \alpha
a_{0}^{\beta _{1}}-6\left( \beta _{1}-2\right) \beta _{4}\rho _{0}\right]
\left( t-t_{0}\right) {}^{2}+...,\nonumber\\
\end{eqnarray}
\bea
H(t)&\approx& \frac{a_{0}^{-\frac{\beta _{1}}{2}}\sqrt{\alpha a_{0}^{\beta _{1}}+12\beta
_{4}\rho _{0}}}{2\sqrt{3}}-\frac{1}{2}\beta _{1}\beta _{4}\rho
_{0}a_{0}^{-\beta _{1}}\times \nonumber\\
&&\left( t-t_{0}\right) +\frac{\beta _{1}^{2}\beta
_{4}\rho _{0}a_{0}^{-\frac{3\beta _{1}}{2}}\sqrt{\alpha a_{0}^{\beta
_{1}}+12\beta _{4}\rho _{0}}}{8\sqrt{3}}\times \nonumber\\
&&\left( t-t_{0}\right)^{2}+...,
\eea
\bea
q(t)&\approx & \frac{6\left( \beta _{1}-2\right) \beta _{4}\rho _{0}-\alpha
a_{0}^{\beta _{1}}}{\alpha a_{0}^{\beta _{1}}+12\beta _{4}\rho _{0}}-\nonumber\\
&&\frac{%
\sqrt{3}\alpha \beta _{1}^{2}a_{0}^{\frac{\beta _{1}}{2}}\left( \beta
_{4}\rho _{0}\right) {}\sqrt{\alpha a_{0}^{\beta _{1}}+12\beta _{4}\rho _{0}}%
}{\left( \alpha a_{0}^{\beta _{1}}+12\beta _{4}\rho _{0}\right) {}^{2}}%
\left( t-t_{0}\right) +\nonumber\\
&&\frac{\alpha \beta _{1}^{3}\beta _{4}\rho _{0}\left(
\alpha a_{0}^{\beta _{1}}-6\beta _{4}\rho _{0}\right) }{4\left( \alpha
a_{0}^{\beta _{1}}+12\beta _{4}\rho _{0}\right) {}^{2}}\left( t-t_{0}\right)
^{2}+...
\eea

These equations describe the main cosmological parameters during the matter dominated era. The expansion is decelerating, and, depending on the model parameters, the deceleration parameter can have a large range of positive values. The transition to the accelerating phase occurs at a time interval $t_{tr}$, which, in the first order approximation is obtained as
\begin{equation}
t_{tr}\approx t_{0}+\frac{a_{0}^{-\frac{\beta _{1}}{2}}\sqrt{\frac{\alpha
a_{0}^{\beta _{1}}}{3\beta _{4}\rho _{0}}+4}\left[ 6\left( \beta
_{1}-2\right) \beta _{4}\rho _{0}-\alpha a_{0}^{\beta _{1}}\right] }{\alpha
\beta _{1}^{2}\sqrt{\beta _{4}\rho _{0}}}.
\end{equation}

Since $t_{tr}$ must be greater than $t_{0}$, it follows that in order for
the model to admit a matter dominated era followed by a transition to an
accelerated phase, the model parameters must satisfy the condition $6\left(
\beta _{1}-2\right) \beta _{4}\rho _{0}-\alpha a_{0}^{\beta _{1}}>0$, or,
equivalently,
\begin{equation}
\left( 1+\frac{3\beta }{2}\right) \frac{\alpha a_{0}^{\beta _{1}}}{6\beta
_{4}\rho _{0}}<1,
\end{equation}%
a condition that can be easiy satisfied by appropriately choosing the free
parameters in the gravitational action.

\section{Discussions and final remarks}\label{sect4}

In the present paper we have considered the framework of the Palatini formalism the gravitational field equations
for the modified gravity $f(R,T)$ theory, implying  a geometry-matter coupling, with the trace of the energy-momentum tensor included as a field variable in the gravitational action. We have derived the field equations by independently varying the metric and the connection in the $f(R,T)$ type gravitational action, and we have formulated them in both the initial metric frame, as well as in the conformal one, in which
the independent connection can be expressed as the Levi-Civita
connection of an auxiliary, energy-momentum trace dependent metric, which is
related to the physical metric by a conformal transformation. Similarly to the metric case \cite{Ha14}
the energy-momentum tensor of the matter is not conserved, and the energy and momentum balance equations take the same form as in the metric theory. Generally, Palatini type theories have  a number of special properties that make them especially attractive
for analyzing strong gravity phenomena, like, for example,  the dynamics of the early Universe or
stellar collapse processes \cite{28a,28b,28c,28d,28e,28f,28f1,28g}. The coupling of the trace of the energy-momentum tensor  with
the curvature scalar generates  some extra-terms in the gravitational field equations, which strongly depend on the
possible functional forms for the geometry - energy momentum trace coupling. If, for example, it would be possible to generate
through the geometry-anergy momentum trace coupling  some repulsive forces, then one
could obtain cosmological models that are non-singular at extremely high densities and
high geometric curvatures, or even one could construct models for non-singular collapsing stars as viable alternatives for the black hole paradigm.

To obtain such repulsive gravitational forces, in modified gravity theories with geometry-matter coupling no new degrees of
freedom in the matter side (exotic sources) or in the gravitational side  are
required in the total action.  In these models the extra-force is
simply induced by the coupling between matter and geometry. Our present results show
that Palatini type theories might play an important role  in the phenomenology
of gravity at both high densities (energies), as well as in the very low
density limit. On the other hand, in the variational process the assumption of independence between
metric and connection  is essential to obtain second
order differential equations for the metric tensor. It is thus possible to assume that at large/small scales
the effective descriptions of the gravitational forces, going beyond standard general relativity,
could come from the Palatini formulation of gravity theories.

In the Palatini type formulation of the $f(R,T)$ gravity, the equation of motion of
massive particles is non-geodesic, and in three dimensions and in the Newtonian limit, Eq. (\ref{force0}) can be formally
represented as an ordinary vector equation in three dimensions of the form $\vec{a} = \vec{a}_N + \vec{a}_f$ where $\vec{a}$ represents
the total acceleration of the particle, $\vec{a}_N$ denotes the Newtonian gravitational
acceleration, while $\vec{a}_f$ is the acceleration  due to the presence
of the extra force induced by the coupling between geometry and matter.  This shows that one observational or experimental possibility of testing the
effects of the coupling between geometry and the trace of the energy-momentum tensor could be in the physical domain of extremely small
particle accelerations, with values of the order of $10^{-10}$ m/s$^2$ . Such an
acceleration could explain the observed behavior of test particles rotating
around galaxies, which is usually explained by postulating the existence of dark matter. However,  as a possible astrophysical application of the gravitational field equations derived with the Palatini formalism  one may consider an alternative view to the dark matter problem, in which the mass discrepancy in galaxies and clusters of galaxies as well as the galactic rotation curves are explained by the existence of a non-minimal coupling between matter and geometry.

We have also briefly investigated the intriguing feature of the non-conservation of the energy - momentum
tensor of the matter in the $f(R,T)$ gravity theory by interpreting it in the framework of the thermodynamics of open systems.
We have interpreted this effect as describing  phenomenologically the particle production in the cosmological fluid filling the Universe, with the extra-terms induced  by the non-minimal coupling between $R$ and $T$ assumed to describe particle creation processes, with the gravitational field acting
as a source for particles. We have explicitly obtain the particle
creation rates, the entropy flux, the creation pressure and entropy
generation rate in a covariant form, as functions of the Lagrangian density $f(R,T)$ of the theory, and of its derivatives, respectively.  On the other hand it is natural to assume that such particle production processes are of the same nature as the similar processes that appear in the framework of the quantum field theory in curved space-times. A static gravitational field does not produce particles. But a time dependent
gravitational field can generate new particles. This interesting analogy between gravitational theories with geometry-matter coupling, and quantum field theory in curved spacetimes may open the possibility of an effective classical description of quantum gravity on small geometric scales.

As a cosmological application of the Palatini formalism of the $f(R,T)$ theory we have briefly considered two classes of cosmological models, corresponding to two choices of the gravitational Lagrange density $f(R,T)=k(R)+g(T)$, with $k(R)\sim R^2$ and $k(R)\sim 1/R$, respectively. In both cases we have assumed for the function $g(T)$ the simple form $g(T)\sim T$. We have explicitly shown that both models can generate accelerating expansion of the Universe, with a power law and an exponential form of the scale factor, respectively.

In both metric and Palatini formulation of $f(R,T)$ gravity, dark energy is interpreted as a material-geometrical fluid, with a negative parameter of the equation of state, for which the function $f(R,T)$ is not known {\it a priori}. Hence, similarly to the case of $f(R)$ gravity, there is a need of a model independent reconstruction of the Lagrangian of theory, which can be done by using some relevant cosmographic techniques \cite{o1,o2} to determine which
$f(R,T)$ model is favored with respect to others.  In the case of $f(R)$ gravity, a cosmographic approach was introduced in \cite{r12}, by assuming that the cosmological principle is valid, and that dark energy can be described as a geometric fluid.  Then, after  expanding the cosmological
observables (the Hubble parameter, the luminosity distance, the apparent magnitude modulus, the effective pressure etc.) into Taylor series, and matching the derivatives of the expansions with cosmological data one can obtain some model independent constraints on the gravitational theory. The coefficients of of the power series of the expansion of the scale factor, calculated at present time (at redshift $z = 0$) are known as the cosmographic series.
The importance of the cosmographic approach is that it does not need the assumption of a specific cosmological model. If the scalar curvature is negligible, the Taylor series of the scale factor around $\Delta t=t-t_0=0$ can be represented as \cite{r12}
\be
\frac{1-a(t)}{H_0}\approx \Delta t+\frac{q_0}{2}H_0\Delta t +\frac{j_0}{6}H_0^2\Delta t^3-\frac{s_0}{24}H_0^3\Delta t^4+....,
\ee
where the jerk parameter $j$ is defined as $j(t)=\left(1/aH^3\right)\left(d^3a/dt^3\right)$, while the snap parameter $s$ is given by $s(t)=\left(1/aH^4\right)\left(d^4a/dt^4\right)$ \cite{r1c}. A strategy to infer the transition redshift $z_{da}$, which indicates the passage of the Universe from a decelerating to an accelerating phase, was proposed, in the framework $f(R)$ gravity, in \cite{r11}. This goal can be achieved by numerically reconstructing $f(z)$, that is, the corresponding gravitational Lagrangian $f(R)$ re-expressed as a function of the redshift $z$, and by matching $f(z)$ with cosmography. The high-redshift $f(R)$ cosmography was considered in \cite{r1c}, by adopting the technique of polynomial reconstruction. Instead of considering the Taylor expansions that proved to be non-predictive for redshifta  $z>1$, the Pad\'e rational approximations were considered, by performing series expansions that converge in the domains of high redshifts. As a first step in this strategy is the reconstruction of the function $f(z)$, by assuming that the Ricci scalar can be inverted with respect to the redshift $z$.

The cosmographic approach developed in \cite{r12,r11,r1c} for the case of $f(R)$ gravity can also be extended to both metric and Palatini $f(R,T)$ gravities. To be more specific, such an approach requires to rewrite $f(R,T)$ or $f\left(R\left(g,\tilde{\Gamma}\right),T\right)$ into a function $f(z)$. Similarly to the approach introduced in  \cite{r1} for $f(R)$ gravity in the metric formulation, in order to handle high-$z$ data one can rewrite the $f(R,T)$ function into an $f(z)$ function generally through the use of the Pad\'e polynomials. As a next step data fitting based on some general $f(z)$ models is required. Hence one can generalize the approaches  of \cite{r12,r11,r1c}, as well as the investigations performed in the framework of general relativity and other Extended Theories of Gravity in \cite{o3,o4,o5} to the case of  the $f(R,T)$ theory, in both metric and Palatini formulations, and numerically determine the coefficients of the series expansions for $R$ and $T$ in the $f(R,T)$ models through cosmological data fitting. The  cosmographic  approach could help distinguish between the roles and weights of the functions $R$ and $T$ in the gravitational action, and lead to a full comparison of the theory with the cosmological observations.

The cosmology of the Palatini $f(R,T)$ gravity can represent a promising way for the explanation of the accelerated phases in the dynamics of the Universe, and which characterized its evolution in both very early and late stages. In the present paper we have introduced some basic theoretical tools necessary for the in depth investigation  of the cosmological and astrophysical aspects  of the Palatini formulation of $f(R,T)$ gravity.

\section*{Acknowledgments}

We would like to thank to the two anonymous reviewers for comments and suggestions that helped us to significantly improve our manuscript. Shi-Dong Liang acknowledges the support of the Natural Science Foundation of
Guangdong Province (grant no. 2016A030313313). T. H. would like to thank the Yat-Sen School of the Sun Yat-Sen University
in Guangzhou, P. R. China, for the kind hospitality offered during the
preparation of this work.
%%%%%%%%%%%%%%%%%%%%%%%%%%%%%%%%%%%%%%%%%%%%%%%%%%%%%%%%%%%%%%%%%%%%%%%%%%%%%%

%%%%%%%%%%%%%%%%%%%%%%%%%%%%%%%%%%%%%%%%%%%%%%%%%%%%%%%%%%%%%%%%%%%%%%%%%%%%%%

\appendix

\section{$f(R,T)$ field equations in the metric formulation\label{app1}}

We first introduce two basic formulae we are going to use,
\begin{equation}
\frac{2 \delta \sqrt{-g}}{\sqrt{-g}} = g^{\mu\nu} \delta g_{\mu\nu} = -
g_{\mu\nu} \delta g^{\mu\nu},
\end{equation}
\begin{eqnarray}
\delta \Gamma_{\thickspace \mu\nu}^{\rho} &=& \frac{1}{2} g^{\rho \lambda}
\cdot 2 g_{\sigma \lambda} \delta \Gamma_{\thickspace \mu\nu}^{\sigma}
\notag \\
&=& \frac{1}{2} g^{\rho \lambda} \left( \nabla_\mu \delta g_{\nu \lambda} +
\nabla_\nu \delta g_{\lambda \mu} - \nabla_\lambda \delta g_{\mu \nu}
\right),  \label{eq:delgamm}
\end{eqnarray}
where in Eq.~\eqref{eq:delgamm} we have taken into account the relations $%
\delta \left( \nabla_\mu g_{\nu \lambda} \right) = \delta \left( \nabla_\nu
g_{\lambda \mu} \right) = \delta \left( \nabla_\lambda g_{\mu\nu} \right) =
0 $ . A variation $\delta g^{\mu\nu}$ of the metric tensor then leads to
%\begin{equation}
%\delta g^{\alpha\beta} = \delta \left( g^{\mu\nu} \delta_\mu^\alpha
%\delta_\nu^\beta \right) = \delta_\mu^\alpha \delta_\nu^\beta \delta
%g^{\mu\nu} = - g^{\alpha \mu} g^{\beta \nu} \delta g_{\mu\nu};
%\end{equation}
\begin{eqnarray}
\delta T &=& \delta \left( g^{\alpha\beta} T_{\alpha\beta} \right) = \left(
T_{\alpha\beta} \frac{\delta g^{\alpha\beta}}{\delta g^{\mu\nu}} +
\Theta_{\mu\nu} \right) \delta g^{\mu\nu}  \notag \\
&=& \left( T_{\mu\nu} + \Theta_{\mu\nu} \right) \delta g^{\mu\nu}, \quad
\Theta_{\mu\nu} \equiv g^{\alpha\beta} \frac{\delta T_{\alpha\beta}}{\delta
g^{\mu\nu}},
\end{eqnarray}
and
\begin{eqnarray}
\delta R_{\mu\nu} &=& \delta \left( \partial_\rho \Gamma_{\thickspace %
\mu\nu}^\rho - \partial_\nu \Gamma_{\thickspace \mu \rho}^\rho + \Gamma_{%
\thickspace \rho \lambda}^\rho \Gamma_{\thickspace \mu\nu}^\lambda - \Gamma_{%
\thickspace \nu \lambda}^\rho \Gamma_{\thickspace \mu \rho}^\lambda \right)
\notag \\
&=& \nabla_\rho \delta \Gamma_{\thickspace \mu\nu}^\rho - \nabla_\nu \delta
\Gamma_{\thickspace \mu \rho}^\rho,
\end{eqnarray}
respectively, where the above relation is called the Palatini identity. Then
for the variation of the metric we obtain
\begin{eqnarray}
\hspace{-0.7cm}&&\delta S = \int \frac{\sqrt{-g}}{16 \pi} \bigl[ f_R \left(
R_{\mu\nu} \delta g^{\mu\nu} + g^{\mu\nu} \delta R_{\mu\nu} \right) + f_T
\delta T  \notag \\
\hspace{-0.5cm}&& - \frac{g_{\mu\nu}}{2} f \delta g^{\mu\nu} - 8 \pi
T_{\mu\nu} \delta g^{\mu\nu} \bigr] \thickspace \mathrm{d}^4 x  \notag \\
\hspace{-0.5cm}&=& \int \frac{\sqrt{-g}}{16 \pi} \bigl[ f_R \left(
R_{\mu\nu} + g_{\mu\nu} \Box - \nabla_\mu \nabla_\nu \right) + f_T \left(
T_{\mu\nu} + \Theta_{\mu\nu} \right)  \notag \\
\hspace{-0.5cm}&& - \frac{g_{\mu\nu}}{2} f - 8 \pi T_{\mu\nu} \bigr] \delta
g^{\mu\nu} \thickspace \mathrm{d}^4 x, \quad \Box \equiv g^{\alpha\beta}
\nabla_\alpha \nabla_\beta.
\end{eqnarray}
Assuming that the variation of $\delta g^{\mu\nu}$ vanishes at infinity,
then
\begin{eqnarray}
\hspace{-0.5cm}&& \int \frac{\sqrt{-g}}{16 \pi} \left( g_{\mu\nu}
g^{\alpha\beta} f_R \nabla_\alpha \nabla_\beta - f_R \nabla_\mu \nabla_\nu
\right) \delta g^{\mu\nu} \thickspace \mathrm{d}^4 x  \notag \\
\hspace{-0.5cm}&=& \int \frac{\sqrt{-g}}{16 \pi} \left( - g_{\mu\nu}
g^{\alpha\beta} \nabla_\alpha f_R \nabla_\beta \delta g^{\mu\nu} +
\nabla_\mu f_R \nabla_\nu \delta g^{\mu\nu} \right) \thickspace \mathrm{d}^4
x  \notag \\
\hspace{-0.5cm}&=& \int \frac{\sqrt{-g}}{16 \pi} \left( g_{\mu\nu}
g^{\alpha\beta} \delta g^{\mu\nu} \nabla_\beta \nabla_\alpha f_R - \delta
g^{\mu\nu} \nabla_\nu \nabla_\mu f_R \right) \thickspace \mathrm{d}^4 x
\notag \\
\hspace{-0.5cm}&=& \int \frac{\sqrt{-g}}{16 \pi} \delta g^{\mu\nu} \left(
g_{\mu\nu} \Box - \nabla_\mu \nabla_\nu \right) f_R \thickspace \mathrm{d}^4
x,
\end{eqnarray}
and
\begin{eqnarray}
\hspace{-1.0cm}&&\delta S = \int \frac{\sqrt{-g}}{16 \pi} \bigl[ \left(
R_{\mu\nu} + g_{\mu\nu} \Box - \nabla_\mu \nabla_\nu \right) f_R +  \notag \\
\hspace{-1.0cm}&& \left( T_{\mu\nu} + \Theta_{\mu\nu} \right) f_T - \frac{%
g_{\mu\nu}}{2} f(R,T) - 8 \pi T_{\mu\nu} \bigr] \delta g^{\mu\nu} %
\thickspace \mathrm{d}^4 x.
\end{eqnarray}

Since $\delta S=0$, from the above relation we obtain immediately the field
equations~\eqref{eq:feqm} of the $f(R,T)$ gravity theory.

\section{Divergence of the matter energy-momentum tensor in the metric
formalism}

\label{app2}

By taking the covariant divergence of Eq.~\eqref{eq:feqm1}, with the use of
the mathematical identity $\nabla_\mu G_{\thickspace\nu}^\mu (g) = 0$ we
obtain
\begin{eqnarray}
&& \nabla_\mu \left[ f_{R} G_{\thickspace\nu}^\mu (g) \right] - R_{\mu\nu}
(g) \nabla^\mu f_{R} + \frac{\delta_{\thickspace\nu}^\mu}{2} R(g) \nabla_\mu
f_{R}  \notag \\
&=& \left[ G_{\thickspace\nu}^\mu (g) - R_{\thickspace\nu}^\mu (g) + \frac{%
\delta_{\thickspace\nu}^\mu}{2} R \right] \nabla_\mu f_{R}  \notag \\
&=& \nabla_\mu \left[ 8 \pi T_{\thickspace\nu}^\mu - \left( T_{\thickspace%
\nu}^\mu + \Theta_{\thickspace\nu}^\mu \right) f_T \right] + \frac{\delta_{%
\thickspace\nu}^\mu}{2} \left( \nabla_\mu f - f_{R} \nabla_\mu R \right)
\notag \\
&& + \left( \Box \nabla_\nu - \nabla_\nu \Box \right) f_{R} - R_{\mu\nu} (g)
\nabla^\mu f_{R}  \notag \\
&=& \nabla_\mu \left[ 8 \pi T_{\thickspace\nu}^\mu - \left( T_{\thickspace%
\nu}^\mu + \Theta_{\thickspace\nu}^\mu \right) f_T \right] + \frac{f_T}{2}
\nabla_\nu T = 0,  \label{eq:covder-Tmunuma}
\end{eqnarray}
where we have used the relations
\begin{eqnarray}
&& \left( \nabla_\nu \Box - \Box \nabla_\nu \right) \phi = g^{\alpha\beta}
\left( \nabla_\nu \nabla_\alpha \nabla_\beta - \nabla_\alpha \nabla_\beta
\nabla_\nu \right) \phi  \notag \\
&=& g^{\alpha\beta} \left( \nabla_\nu \nabla_\alpha - \nabla_\alpha
\nabla_\nu \right) \nabla_\beta \phi = g^{\alpha \beta} R_{\thickspace \beta
\alpha \nu}^\mu \nabla_\mu \phi  \notag \\
&=& - R_{\mu \nu} \nabla^\mu \phi,
\end{eqnarray}
and
\begin{equation}
\nabla_\nu f \left( {R}, T \right) = f_{R} \nabla_\nu R + f_T \nabla_\nu T,
\label{eq:nablaf}
\end{equation}
respectively.

\section{The geometric quantities in the FRW geometry}

\label{app3}

For the metric~\eqref{eq:FRW-metric-flat}, we have
\begin{equation}
\partial_\lambda g^{\mu\nu} = \left\{
\begin{array}{rl}
2 a \dot a, & \lambda = 0 \text{ and } \mu = \nu \neq 0 \\
0, & \text{others}%
\end{array}
\right..
\end{equation}
Hence the only nonzero components of the connection, $\Gamma_{\thickspace %
\mu\nu}^{\rho} (g) = \left(g^{\rho \sigma}/2\right) \left( \partial_\nu
g_{\sigma\mu} + \partial_\mu g_{\sigma\nu} - \partial_\sigma g_{\mu\nu}
\right)$, are now
\begin{equation}
\Gamma_{\thickspace ii}^0 = a \dot a = a^2 H \text{ and } \Gamma_{%
\thickspace 0i}^i = \Gamma_{\thickspace i0}^i = \frac{\dot a}{a} = H, \quad
i = 1, 2, 3.  \label{eq:non0-cons}
\end{equation}
In that case,
\begin{eqnarray}  \label{R00}
R_{00} (g) &=& \partial_\rho \Gamma_{\thickspace 00}^\rho - \partial_0
\Gamma_{\thickspace 0 \rho}^\rho + \Gamma_{\thickspace \sigma \rho}^\rho
\Gamma_{\thickspace 00}^\sigma - \Gamma_{\thickspace 0 \sigma}^\rho \Gamma_{%
\thickspace 0 \rho}^\sigma  \notag \\
&=& - 3 \left( \dot H + H^2 \right),
\end{eqnarray}
\begin{equation}  \label{Rii}
R_{ii}(g) = a^2 \left( \dot H + 3 H^2 \right),
\end{equation}
and
\begin{eqnarray}  \label{eq:R(H)}
R (g) &=& g^{\mu\nu} R_{\mu\nu} (g) = g^{00} R_{00} (g) + 3 g^{11} R_{11}
(g) =  \notag \\
&& 6 \left( \dot H + 2 H^2 \right).
\end{eqnarray}
Hence
\begin{equation}  \label{A6}
G_{00} (g) = R_{00} (g) -\frac{g_{00}}{2} R (g) = 3 H^2,
\end{equation}
\begin{equation}  \label{A7}
G_{ii} (g) = R_{ii} (g) -\frac{g_{ii}}{2} R (g) = - a^2 \left( 2 \dot{H} + 3
H^2 \right),
\end{equation}

Substituting Eqs.~\eqref{R00} and \eqref{Rii} into Eq.~\eqref{rictenc} we
obtain
\begin{eqnarray}
\hspace{-0.5cm}&& \tilde{R}_{00} \left(\tilde{\Gamma}\right) = R_{00} (g) +
\frac{1}{F} \left[ \frac{3 \left( \nabla_0 F \right)^2}{2 F} - \left(
\nabla_0 \nabla_0 + \frac{g_{00}}{2} \Box \right) F \right]  \notag \\
\hspace{-0.5cm}&=& - 3 \left( \dot H + H^2 \right) + \frac{1}{F} \left[
\frac{3 {\dot{F}}^2}{2 F} - \left( \ddot F + \frac{1}{2} \ddot F + \frac{3H}{%
2} \dot F \right) \right]  \notag \\
\hspace{-0.5cm}&=& - 3 \left( \dot H + H^2 \right) + \frac{3}{2 F} \left(
\frac{ {\dot{F}}^2}{F} - \ddot F - H \dot F \right)  \label{eq:Riccitensor2}
\end{eqnarray}
and
\begin{eqnarray}
&& \tilde{R}_{ii} \left(\tilde{\Gamma}\right) = R_{ii} (g) + \frac{1}{F} %
\left[ \frac{3 \left( \nabla_i F \right)^2}{2 F} - \left( \nabla_i \nabla_i
+ \frac{g_{ii}}{2} \Box \right) F \right]  \notag \\
&=& a^2 \left( \dot H + 3 H^2 \right) + \frac{a^2}{F} \left(H \dot F + H
\dot F\frac{\ddot F}{2} + \frac{3H \dot F}{2} \right)  \notag \\
&=& a^2 \left[ \left( \dot H + 3 H^2 \right) + \frac{1}{2F} \left(\ddot F +
5H \dot F \right) \right].  \label{eq:Riccitensor3}
\end{eqnarray}
Similarly, by substituting Eq.~\eqref{eq:R(H)} into Eq.~\eqref{ricscalc} we
find
\begin{eqnarray}
&& R \left(g,\tilde{\Gamma}\right) = R (g) + \frac{3}{F} \left[ \frac{
\left( \nabla F \right)^2}{2 F} - \Box F \right]  \notag \\
&=& 6 \left( \dot H + 2 H^2 \right) - \frac{3}{F} \left( \frac{{\dot{F}}^2}{%
2 F} - \ddot F - 3 H \dot F \right).  \label{eq:Ricciscalar3}
\end{eqnarray}

Some combinations of the above equations lead to
\begin{eqnarray}  \label{G00}
&& G_{00} \left(g,\tilde{\Gamma}\right) = \tilde{R}_{00} \left(\tilde{\Gamma}%
\right) - \frac{g_{00}}{2} R\left(g,\tilde{\Gamma}\right)  \notag \\
&=& G_{00}(g) + \frac{1}{F} \bigg\{ \left( g_{00} \Box - \nabla_{0}
\nabla_{0} \right) F  \notag \\
&& + \frac{3}{2F} \Big[ \nabla_{0} F \; \nabla_{0}F - \frac{g_{00}}{2}
\left(\nabla F \right)^2 \Big] \bigg\}  \notag \\
&=& 3 H^2 + \frac{1}{2F} \left( \frac{ 3 {\dot{F}}^2}{2F} + 6 H \dot F
\right)
\end{eqnarray}
and
\begin{eqnarray}  \label{Gii}
&& G_{ii} \left(g,\tilde{\Gamma}\right) = \tilde{R}_{ii} \left(\tilde{\Gamma}%
\right) - \frac{g_{ii}}{2} R\left(g,\tilde{\Gamma}\right)  \notag \\
&=& G_{ii}(g) + \frac{1}{F} \bigg\{ \left( g_{ii} \Box - \nabla_{i}
\nabla_{j} \right) F  \notag \\
&& + \frac{3}{2F} \Big[ \nabla_{i} F \; \nabla_{i} F - \frac{g_{ii}}{2}
\left(\nabla F \right)^2 \Big] \bigg\}  \notag \\
&=& a^2 \left[ - \left( 2 \dot{H} + 3 H^2 \right) + \frac{1}{2F} \left(
\frac{ 3 {\dot{F}}^2}{2F} - 2 \ddot F - 4 H \dot F \right) \right].  \notag
\\
&&
\end{eqnarray}
Note that  for the above calculations we have used the relations
\begin{equation}  \label{A13}
\nabla_\mu F = \partial_\mu F = \left(
\begin{array}{cccc}
\dot F, & 0, & 0, & 0,%
\end{array}
\right),
\end{equation}
\begin{equation}  \label{A14}
\nabla_0 \nabla_0 F = \left( \partial_{00} - \Gamma_{\thickspace 00}^\lambda
\partial_\lambda \right) F = \partial_{00} F = \ddot F,
\end{equation}
\begin{equation}  \label{A15}
\nabla_i \nabla_i F = \left( \partial_{ii} - \Gamma_{\thickspace ii}^\lambda
\partial_\lambda \right) F = - \Gamma_{\thickspace ii}^0 \partial_0 F = -
a^2 H \dot F,
\end{equation}
and
\begin{eqnarray}  \label{A16}
\Box F &=& g^{\mu\nu} \nabla_{\mu\nu} F = g^{\mu\nu} \left( \partial_\mu
\partial_\nu - \Gamma_{\thickspace \mu\nu}^\lambda \partial_\lambda \right) F
\notag \\
&=& \left( - \partial_{00} - 3 g^{11} \Gamma_{\thickspace 11}^0 \partial_0
\right) F = - \ddot F - \frac{3 \dot a}{a} \dot F  \notag \\
&=& - \ddot F - 3 H \dot F,
\end{eqnarray}
respectively.

\end{document}